\documentclass[aps,prb,twocolumn,superscriptaddress,groupedaddress]{revtex4}
\usepackage{hyperref}
\usepackage{amssymb}
\usepackage{graphicx}
\usepackage{amsmath}
\usepackage{subfigure}
\usepackage{sidecap}
\usepackage{amsfonts}
\usepackage{xcolor}

\begin{document}
\title{Phases of  frustrated quantum antiferromagnets on the square and triangular lattices}
\author{Yue Yu and Steven~A.~Kivelson}
\affiliation{Department of Physics, Stanford University, Stanford, CA 94305, USA}

\begin{abstract}
We analyze the zero temperature phase diagrams of the spin $S$ quantum antiferromagnet on square and triangular lattices with competing nearest and next-nearest exchange interactions as well as biquadratic couplings.  We approach the problem from the large $S$ limit.  Our primary focus is on determining the extent to which the existence and character of any quantum disordered phases  can be inferred from this approach.
\end{abstract}

\maketitle

\section{Introduction}
A host of interesting quantum disordered phases -- including various flavors of quantum spin liquids, valence-bond solids, or quantum nematics  -- have by now been shown to exist as a matter of principle.\cite{broholm2020quantum,Fawang,chayes1989valence,xu2008ising,fernandes2016vestigial,fernandes2014drives,chubukov2015origin}  However, the issue of where they exist in the $T=0$ phase diagram of simple models of quantum antiferromagnets with microscopically plausible interactions is still incompletely understood.  Various numerical and other approaches have provided strong evidence\cite{jiang2009phase,jiang2012spin,figueirido1990exact,richter2015spin,gong2014plaquette,capriotti2001resonating,iqbal2016intertwined,hu2013direct,li2012gapped,morita2015quantum,wang2016tensor,iqbal2016spin,hu2015competing,kaneko2014gapless,zhu2015spin} that there is a narrow quantum disordered regime in spin $S=1/2$ and even $S=1$ antiferromagnets near the point at which the classical ($S\to \infty$) model would undergo a transition from an ordered state favored by the nearest-neighbor interaction $J_1$ and that favored when the second-neighbor interaction, $J_2$ is sufficiently large.\cite{chandracolemanlarkin,zhong1993spin,chubukov1992order}

With this physics in mind, we have  studied the ground-state  phase diagrams of a family of frustrated quantum antiferromagnets on the 2D square and triangular lattices.    The classical  $S\to \infty$ limit is readily analyzed, and indeed for all the models considered here, this analysis has been carried out previously.\cite{Hayden2010}  We have extended these results by computing the leading order  (and in some cases higher order) corrections to various quantities in powers of $1/S$.  Much of this analysis has been carried out previously as well.\cite{premi,Fang2008,luo2016spin,stanek2011self,ye2017quantum,chernyshev2009spin}  However, by extending the class of models we have considered, and by taking seriously into account the asymptotic character of the $1/S$ expansion,\cite{sudip} we have managed to obtain a fuller, and more readily justified picture of the phase diagrams.  Since for the classical pure $J_1-J_2$ model, the transition point at $J_2=\frac 1 2 J_1$ is highly multi-critical, we have augmented the model studied by including a weak-biquadratic interaction.
%Our primary focus is on identifying and characterizing  any quantum disordered phases  that can be inferred from this approach, given that such phases are  highly quantum mechanical in nature.

\begin{figure}%[htbp][hbt]
	[ht]
	\centering
	\subfigure[$K>0$ square lattice]
	{\includegraphics[width=4cm]{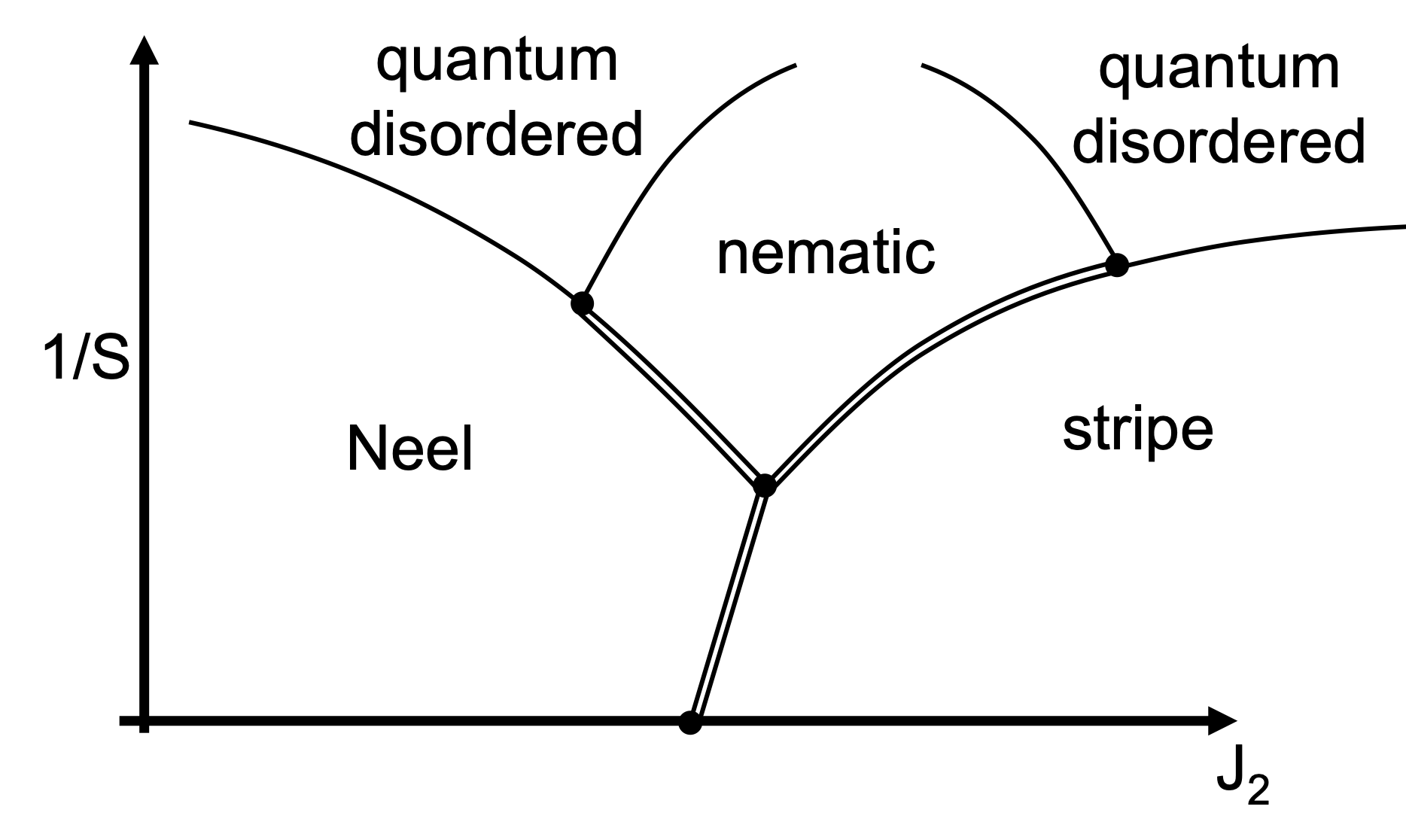} 
		\label{1a}}
	\subfigure[$K<0$ square lattice]
	{\includegraphics[width=4cm]{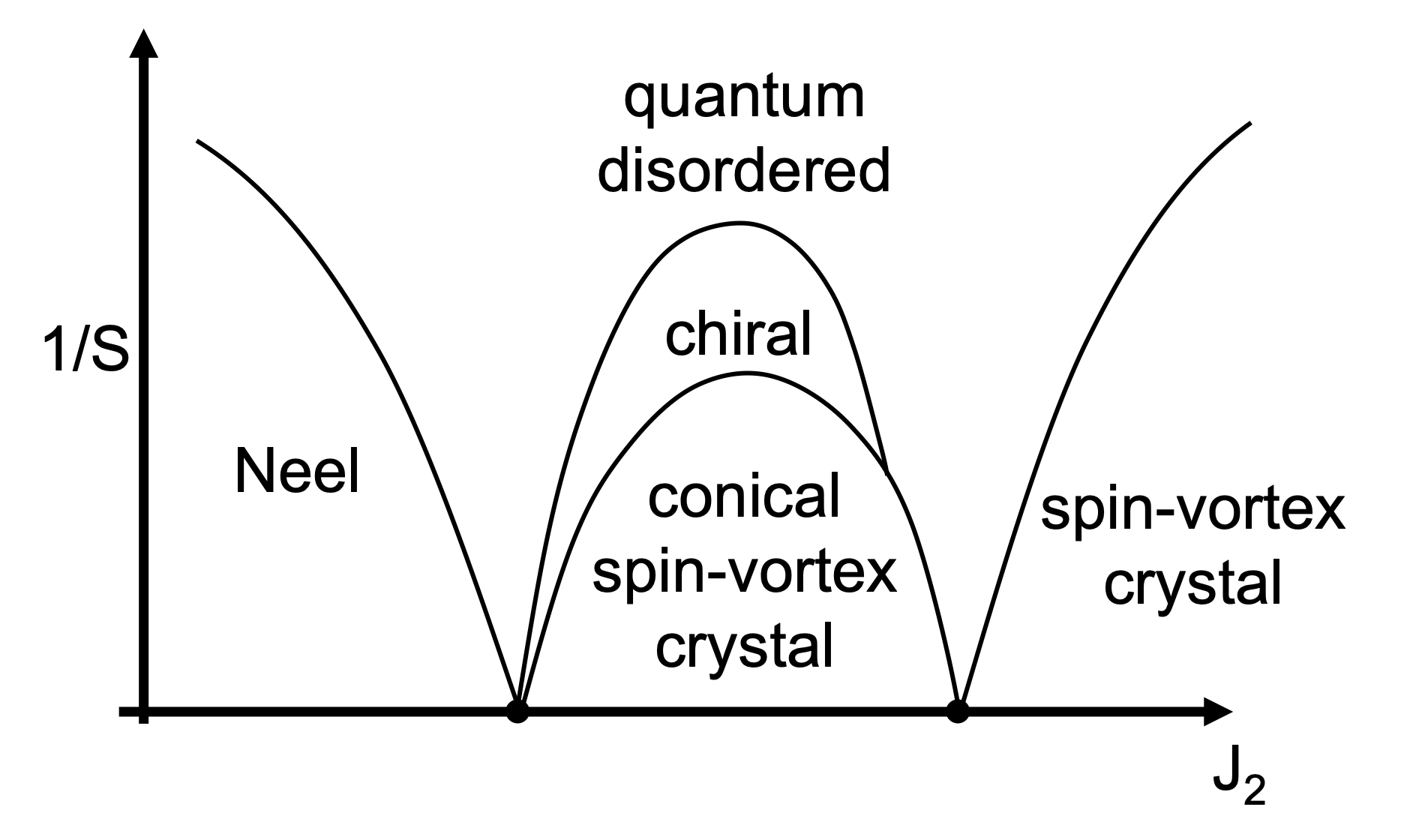}
		\label{1b}}
	\subfigure[$K>0$ triangular lattice]
	{\includegraphics[width=4cm]{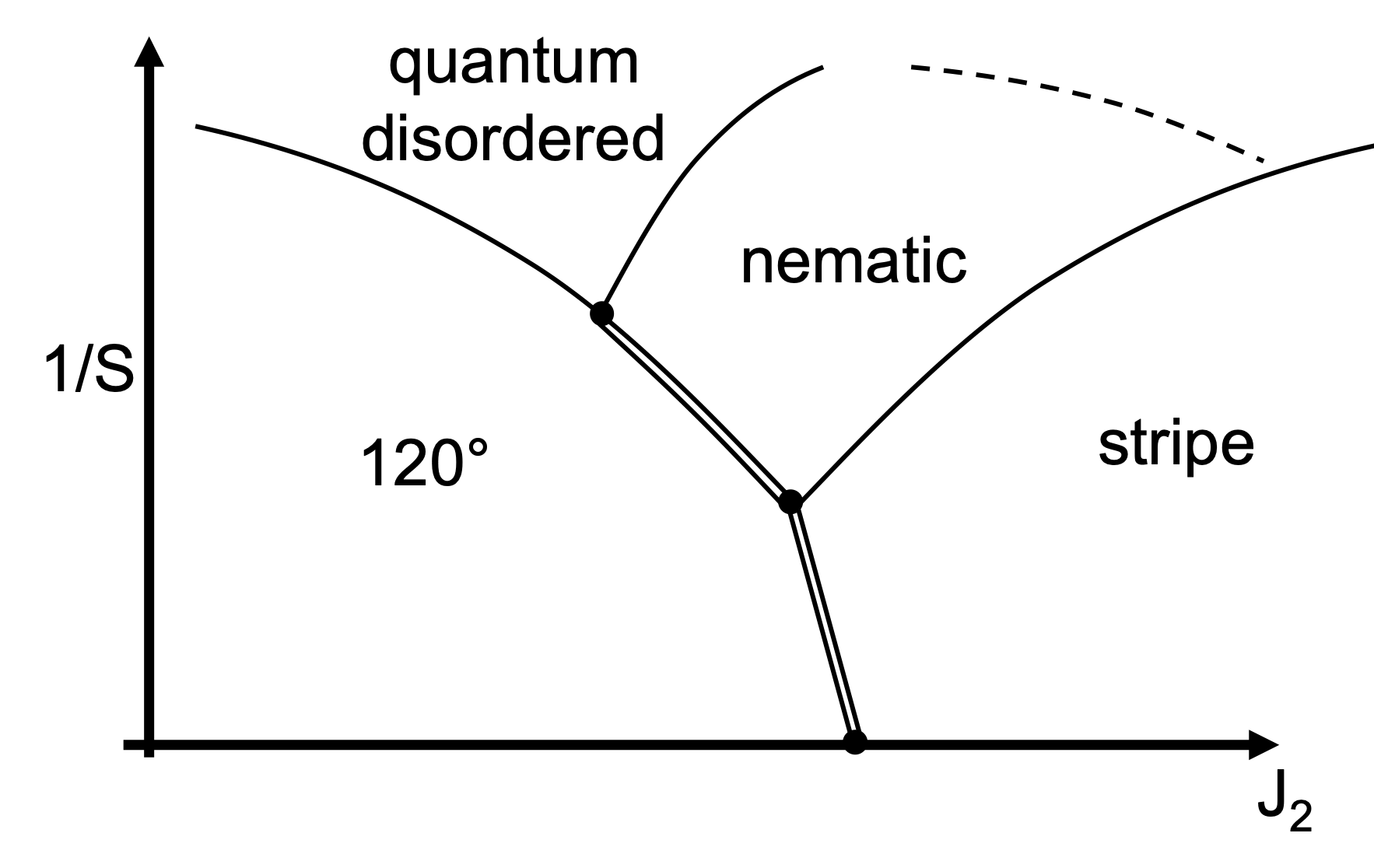} 
		\label{1c}}
	\subfigure[$K<0$ triangular lattice]
	{\includegraphics[width=4cm]{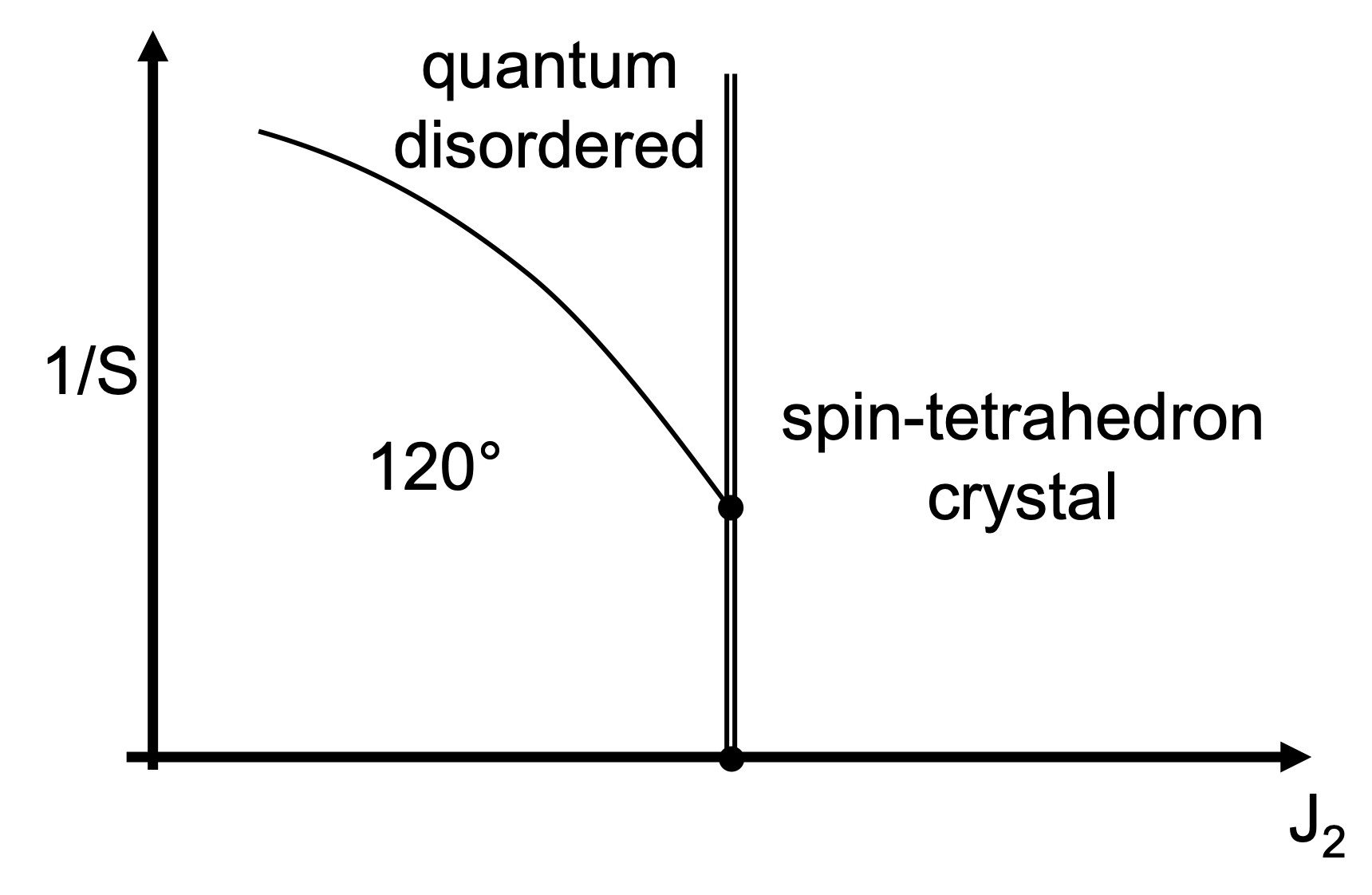}
		\label{1d}}
	\caption{ Schematic ground-state phase diagrams for the square lattice spin $S$ antiferromagnet with $K>0$ and $K<0$ are shown in panels a and b, respectively. Panels c and d are for the triangular lattice with $K>0$ and $K<0$ respectively.  The large $S$ portions of the phase diagrams follow directly from the present analysis - the  small $S$ portions involve extrapolation and plausibility arguments.  In the ``nematic,'' ``chiral,'' and ``quantum disordered'' phases, quantum fluctuations are sufficient to destroy magnetic long-range order but in the first two of these we present suggestive evidence that vestigial order of the indicated variety survives from the nearby ordered phases.  Moreover, in the quantum disordered phases, additional forms of order, including valence-bond crystalline and topological order may arise in ways that depend crucially on whether $S$ is  even or odd integer or half-integer.
	}
	\label{schematic}
\end{figure}

Our principle results, as we will discuss, are summarized in the  schematic phase diagrams shown in Figs. \ref{schematic}.  %There are various places in the generalized phase diagram where various studies (especially DMRG and variational) on the model with S=1/2 or even S=1 suggest that quantum disordered phases -- quantum spin liquids, valence-bond-solids, or quantum nematics -- arise in a narrow  range of parameters between two magnetically ordered phase.  
 For the most part at large $S$ (where our results are most reliable), instead of an intermediate quantum disordered phase, we find %instead 
 direct first order transitions, for instance between a Neel and stripe phase on the square lattice or the three-sublattice 120$^o$ antiferromagnet and the stripe phase on the triangular lattice.  
However, an exponentially narrow regime of a quantum disordered phase appears on the square lattice between the spin-vortex crystal (SVC)  and  the conical spin-vortex crystal (CSVC) phases (described in Fig. \ref{classical}) and between the CSVC  and the Neel phase.
On the other hand, if we extrapolate our results to smaller $S$, we find evidence for  regimes of quantum disordered phases in the same regime of couplings suggested by earlier numerical studies, as also shown in the figure.  %We  find evidence of a bicritical point,  .....
%However, 
Not reflected in the figure are a number of  subtleties -- including the effects of specific topological considerations associated with the quantization of $S$ -- that can alter the nature of the transitions and of the various quantum disordered phases  in the phase diagrams in Fig. \ref{schematic};  we discuss some of these subtleties and other ambiguities in Sec. \ref{subtleties} below.

%***There are various places in which quantum disordered phases appear in these phase diagrams.  However, in all cases they occur in relatively narrow (fine tuned) ranges of interactions so long as $S$ is large, and never in regimes in which the $1/S$ expansion is entirely reliable.  Extrapolating our results to the smallest possible spins, $S=1$ or $S=1/2$, is of course dangerous, but at least provides suggestive guidance for future numerical studies in terms of where in parameter space such phases are most likely to arise, and what their character may be (i.e. what forms of vestigial order\cite{nie,dunghaiandme} they may exhibit).

\section{The model}
We considered interacting spin $S$ operators on  a regular $2D$ lattice, with nearest and next-nearest-neighbor quadratic interactions as well as nearest-neighbor biquadratic interactions
\begin{eqnarray}
H=&& \frac {%J_1
1}{S^2}\sum_{\langle{ij}\rangle}\vec{S_i}\cdot\vec{S_j}+\frac{J_2}{S^2}\sum_{\langle\langle{ik}\rangle\rangle}\vec{S_i}\cdot\vec{S_k} \\
&&-\frac {K}{S^4}\sum_{\langle{ij}\rangle}(\vec{S_i}\cdot\vec{S_j})^2 - \frac {K^\prime} {S^4} \sum_{\langle ijkl\rangle} (\vec S_i\cdot\vec S_j)(\vec S_k\cdot \vec S_l) \nonumber
\end{eqnarray}
where $\vec S_j$ satisfy canonical commutation relations for spin operators, $[S_{j}^a,S_k^b] = i\delta_{ij} \epsilon^{abc} S_j^c$ with $\vec S_j\cdot \vec S_j =S(S+1)$.
Since  we will always assume that the nearest-neighbor exchange coupling, $J_1>0$, we can chose units of energy such that $J_1=1$ as in the above. We will consider explicitly the cases of a square and a triangular lattice.  Here $\langle ij\rangle$ and $\langle\langle ij \rangle\rangle$ denote nearest and next-nearest neighbors respectively and $\langle ijkl\rangle$ signifies sites forming minimal squares on the lattice.
We have normalized the interactions so that the ground-state energy density has a well defined $S \to \infty$ limit, when the spins can be treated as classical Heisenberg rotors.  

For large $S$ the effects of $K$ and $K^\prime$ are very similar.  For the most part, we will report explicit results for $K^\prime=0$, as this is slightly more convenient for the large $S$ analysis.  However, there is a significant difference for $S=1/2$, where $K$ can be incorporated exactly into a renormalized value of $J_1$, while $K^\prime$ remains an independent coupling constant.  Thus, when extrapolating our results to  $S=1/2$, one should loosely interpret $K$ as a proxy for $K^\prime$.

\section{Classical phase diagrams}
\subsection{ The square lattice}
\begin{figure}[h]
\includegraphics[width=8cm]{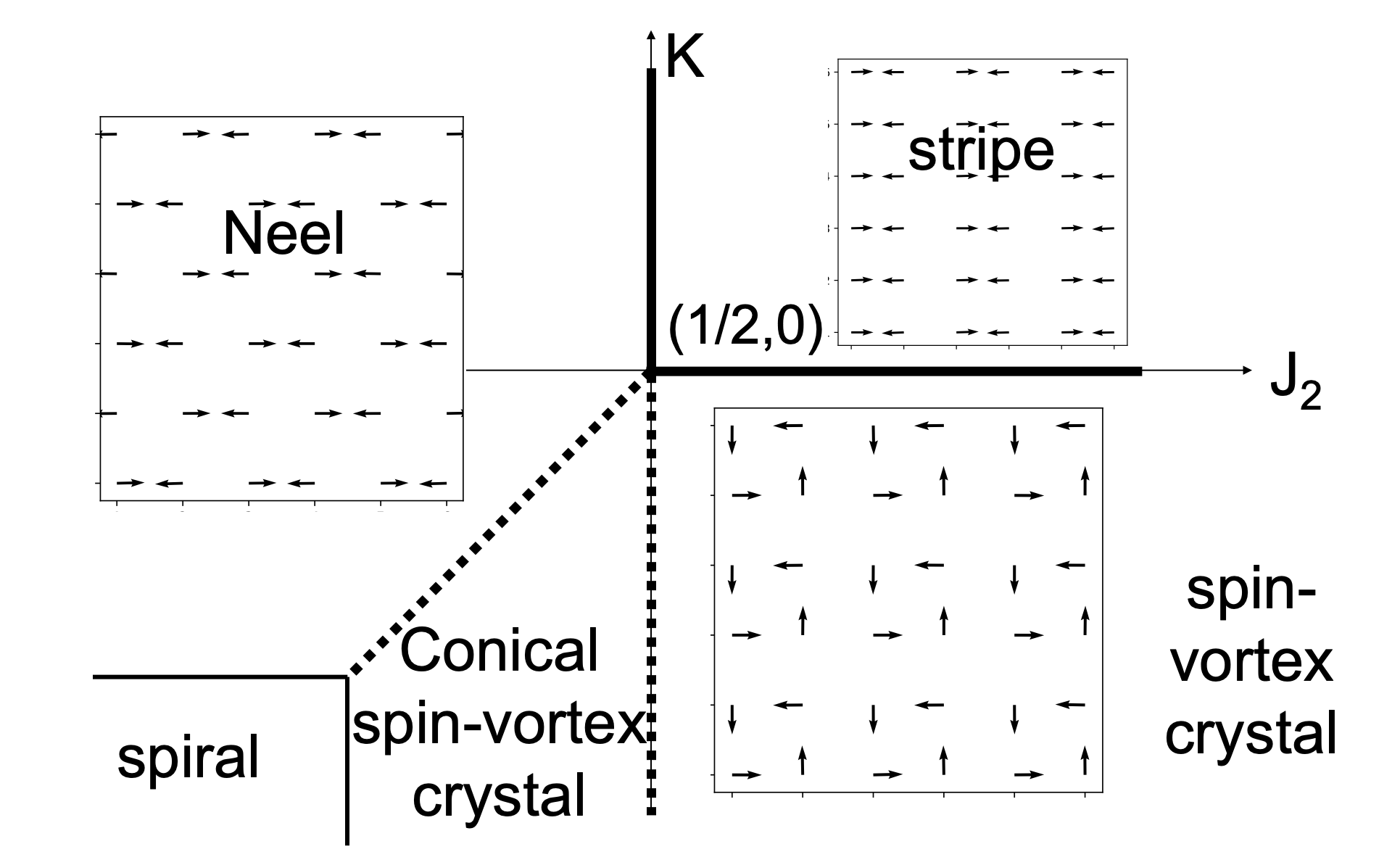}
\caption{Classical phase diagram for the square lattice. Collinear Neel and stripe phases are preferred by positive $K$. %Non-collinear 
The coplanar (non-colinear) spin-vortex-crystal (SVC) and non-coplanar conical spin-vortex-crystal (CSVC) phases are preferred by negative $K$. 
In the CSVC  the (spontaneously chosen) XY-components of the spins order as in the SVC, and  the z-components exhibit Neel order. } 
\label{classical}
\end{figure}

The zero temperature classical phase diagram in the $J_2-K$ plane for square lattice was discussed in Ref. \onlinecite{Hayden2010}, and is summarized in Figs.\ref{classical}.

When $K>0$, there is a first order phase transition between the collinear Neel and the collinear stripe phases at $J_2=J_1/2$.
By contrast, for $K<0$ and of small magnitude, there are three phases as a function of increasing $J_2$:  a Neel phase, a non-coplanar conical spin-vortex crystal (CSVC) phase,  and a coplanar (non-collinear) spin-vortex crystal (SVC) phase. 
 The SVC state,  illustrated in Fig.\ref{classical}, consists of alternating spin-vortices on neighboring plaquettes, oriented with respect to a spontaneously chosen ``X-Y''  plane in spin-space.
 The CSVC can be thought as a linear combination of the Neel %state 
 and SVC states, with spin-components  in the preferred X-Y plane  oriented as in a SVC state, while  the z components exhibit Neel-type order.  It can thus be viewed as a state with coexisting Neel and SVC order, and correspondingly the
 two phase transitions at $J_2=J_1/2+K$ and $J_2=J_1/2$ are continuous. 
As we will focus only on reasonably small $|K|$, we will  neglect the spiral phase that arises when $K$ is large and negative. 

Note that $K=0$ is a non-generic line along which the classical ground states at $J_2>J_1/2$ are highly degenerate. 
%The degeneracy can be lifted through order by disorder phenomenon when quantum ($1/S$) corrections % is
%are  taken into consideration. %The resulted state is the stripe state, which suffers less from quantum fluctuation. 
%
%In this work, we will study the effect of the nearest neighbor biquadratic interaction under first order quantum fluctuation, using linear spin wave theory. At classical level, non-zero biquadratic interaction will lift the degeneracy according to Fig.\ref{classical}. An attractive interaction will further enhance the ordering in Neel and stripe order, alongside with the order by disorder phenomenon. We calculated the ground energy under first order quantum correction, and found that the first order phase transition between Neel and stripe phase is robust at large spin.
Here the $1/S$ analysis is subtle, involving effects of ``order from disorder.''

\subsection{ The triangular lattice}
\begin{figure}[h]
\centering
\includegraphics[width=8cm]{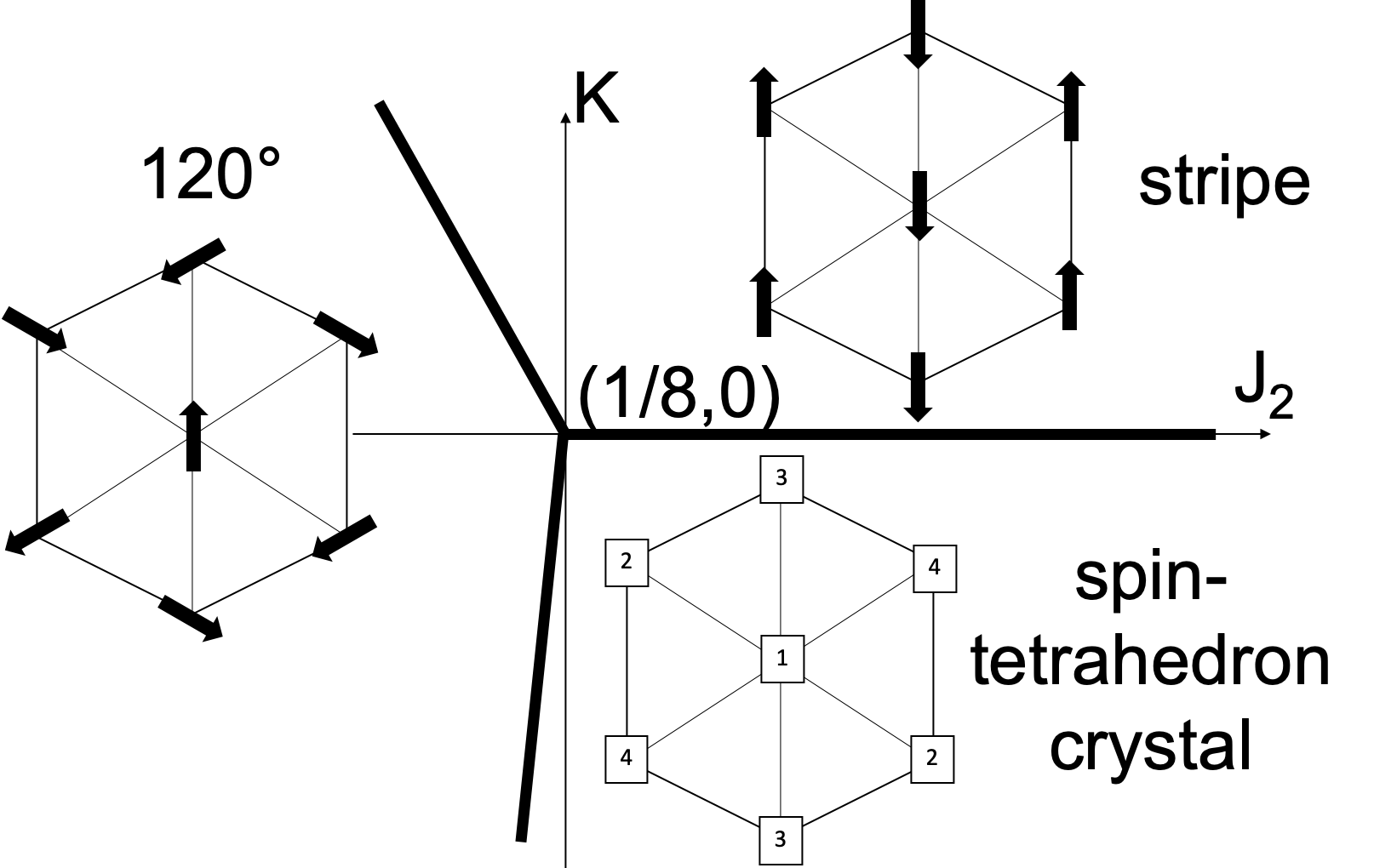}
\caption{Classical phase diagram for the triangular lattice. The $120^\circ$ phase has 3-sublattice coplanar spin order, and the stripe phase has 2-sublattice colinear spin order, as shown in the figure. The spin-tetrahedron-crystal (STC) phase has %4-sublattice 3D 
non-coplanar order in which the spins on the four sublattices point to distinct vertices of a tetrahedron.}
\label{Tf1}
\end{figure}

The zero temperature classical phase diagram in the $J_2-K$ plane for the triangular lattice  is summarized in Fig.\ref{Tf1}.
For $K>0$ there is a first order phase transition from a $120^\circ$ three-sublattice phase for $J_2<\frac{J_1}{8}-\frac{9K}{16}$ to a two-subblattice stripe phase for larger $J_2$. For $K<0$, the system undergoes a first order phase transition from the $120^\circ$ phase to a four-sublattice non-coplanar spin-tetrahedron crystal (STC) phase at a critical value of $J_2=\frac{J_1}{8}+\frac{5K}{48}$.

The $120^\circ$ and stripe phases are shown schematically in Fig.\ref{Tf1}.
The STC %can be decomposed into 
has a 4 sublattice structure, as also shown, such that the four spins (up to a global rotation) point in the direction of the four vertices of a tetrahedron. It is easily checked that this configuration minimizes the repulsive biquadratic interaction in the 4-sublattice decomposition. 
We will restrict attention to relatively small biquadratic interactions $K$, and small $|J_2-J_1/8|$; other interesting phases could appear for stronger interactions. %{\color{red} we also need small $|J_2-J_1/8|$}

\section{First order ($1/S$) quantum corrections}
A systematic formalism for computing quantum fluctuations about the classical ground-state can be accomplished using a Holstein-Primikoff (HP) transformation to map the problem into a problem of weakly interacting bosons. The first order corrections in powers of $1/S$ are obtained by keeping terms to quadratic order in the bosonic fields - i.e. treating the quantum fluctuations as non-interacting spin-waves. Higher order corrections in powers of $1/S$ can be computed, in principle, by treating the interactions between bosons perturbatively.   
This formalism, and the details of specific calculations are reviewed in the appendix. In this section, we report on the quantum corrections to various quantities computed to first order in $1/S$.

The ground-state energy per lattice site of a system in a given state labeled by $a$ (for example, $a=$Neel) is 
\begin{eqnarray}
E_a=&&E_{cl,a}+\frac 1 {SN}\left[\sum_{k}\omega_{{\bf k},a}+C_{K,a}\right] \ldots \\
=&& E_{cl,a}+S^{-1} {\cal E}_a+ \ldots  \nonumber
\end{eqnarray}    
where $E_{cl,a}$ is the classical ground-state energy, $N$ is the number of sites, $\omega_{{\bf k},a}$ are the normal mode frequencies,  ${\bf k}$ (which specifies the Bloch wave-number and possibly other quantum numbers where necessary) labels the individual normal modes, $C_{K,a}$ is a constant term proportional to K from $\vec{S_i}\cdot\vec{S_i}=S(S+1)$ terms in various of non-collinear states, and $\ldots$ indicates higher order terms in powers of $1/S$.  In particular, the non-interacting part of the HP Hamiltonian can be expressed in terms of bosonic creation operators, $b_{\bf k}^\dagger$, as $H_0 = \sum_{\bf k}\left[A_{\bf k} b_{\bf k}^\dagger b_{\bf k} + B_{\bf k}b_{-{\bf k}}^\dagger b_{\bf k}^\dagger + {\rm H.C.}\right]$ where the coefficients $A_{\bf k}$ and $B_{\bf k}$ depend on the nature of the classical ordered state that serves as the starting point, and 
\begin{equation}
\omega_{{\bf k},a}=\sqrt{|A_{\bf k}|^2-|B_{\bf k}|^2}-A_{\bf k} \ .
 \end{equation}
 Manifestly, this calculation only makes sense so long as $\omega_{{\bf k},a}$ is real for all ${\bf k}$, i.e. that $|A_{\bf k}|^2-|B_{\bf k}|^2\geq{0}$.
This is equivalent to the condition that the classical configuration be at least metastable. 
The specifics of the calculations of $E_{a,cl}$ and ${\cal E}_a$ for each of the relevant states are discussed in the Appendix.

Similarly, we will compute the anomalous expectation values of the various order parameters that characterize aspects of the broken symmetries of the various phases:
\begin{equation}
\left |\langle {{\bf O}}_a \rangle \right|=1 - {\cal S}_a S^{-1} + \ldots 
\label{OP}
\end{equation}
where we will always normalize ${\bf O}_a$ so that the classical expectation value of its magnitude is 1, and ${\cal S}_a$ is a function of $J_2$ and $K$. %{\color{red} Changes are made: ${\cal S}_a$ does not have to be positive, for example in nematic phase, it could be negative}
 In particular, the sublattice magnetization of the various phases is defined as 
\begin{eqnarray}
&&m_a\equiv\ %\overline
\frac 1 {SN} \sum_{\bf R}{\langle{\vec S}_{\bf{R}}\cdot{\hat{n}^{cl}_{\bf R}}\rangle}%=1-{\cal S}_a\ S^{-1} + \ldots 
\label{mag}
\end{eqnarray}
where  $\hat{n}^{cl}_{\bf R}$ is a unit vector in the direction (in spin-space) of the corresponding classical orientation of spin at site ${\bf R}$. Again, the explicit calculations pertinent to computing ${\cal S}_a$  are summarized  in the Appendix.

Needless to say, this expansion is strictly justified only when the corrections to the classical results are small.  
Where there is a first order transition, the order parameter is generally non-zero even %at the %critical point
proximate to the transition.  We can determine the location of such phase boundaries by comparing the energy per site, $E_a=E_{a^\prime}$, of the two relevant phases. Thus, the first order corrections to the position of such a phase boundary (e.g. the critical value of $J_2$) can be directly computed from the first order expression for the ground-state energies.  

We will also makes {\it estimates} of the points at which quantum fluctuations become sufficiently large that they cause a classical order parameter to vanish.  Here we are always extrapolating our results beyond the range in which the spin-wave expansion is controlled.  Moreover, since the $1/S$ expansion is known to be an asymptotic series, %\cite{spinwave}, 
 there is no reason to think that this estimate would be improved by keeping higher order terms in the expansion (at least without employing additional information to allow a resumation of the series).  
None-the-less, it is instructive to extrapolate the results to the point at which the leading order expression for each order parameter would vanish;  in this way, we interpret ${\cal S}_a$ as an estimate of the critical value of $S_{crit} \approx {\cal S}_a$ at which each of these order parameters would vanish (baring any other preemptive phase transition that destroys the order at larger $S$).

\subsubsection{Square lattice with $K>0$}
The two pertinent phases for $K>0$ are the Neel and the stripe phase.  Both are states with non-zero sublattice magnetization, $m_a$, with  $a=$Neel and $a=$str (for  stripe order).
The stripe phase also breaks the lattice 4-fold rotational symmetry in a manner that is characterized by the nematic order parameter,
\begin{eqnarray}
&&
\langle {\bf O}_{nem}\rangle =%\equiv 
\frac 1 {2NS^2}\ %\overline
\sum_{\vec R}{\langle{\vec S}_{\bf{R}}\cdot\vec S_{\bf{R+x}}-{\vec S}_{\bf{R}}\cdot\vec S_{\bf{R+y}}\rangle}. %2S(S-\beta)
\end{eqnarray}
The factor of $1/2NS^2$ is included so that in the classical stripe-ordered state, $|\langle {\bf O}_{nem}\rangle |=1$, so that ${\cal S}_{nem}$ is defined as in Eq. \ref{OP}.

The first order quantum corrections to all these orders were computed previously.\cite{Fang2008} We have added to these results expressions for the first order shifts in the ground-state energies, ${\cal E}_a$.
  
\begin{figure}[h]
\includegraphics[width=8cm]{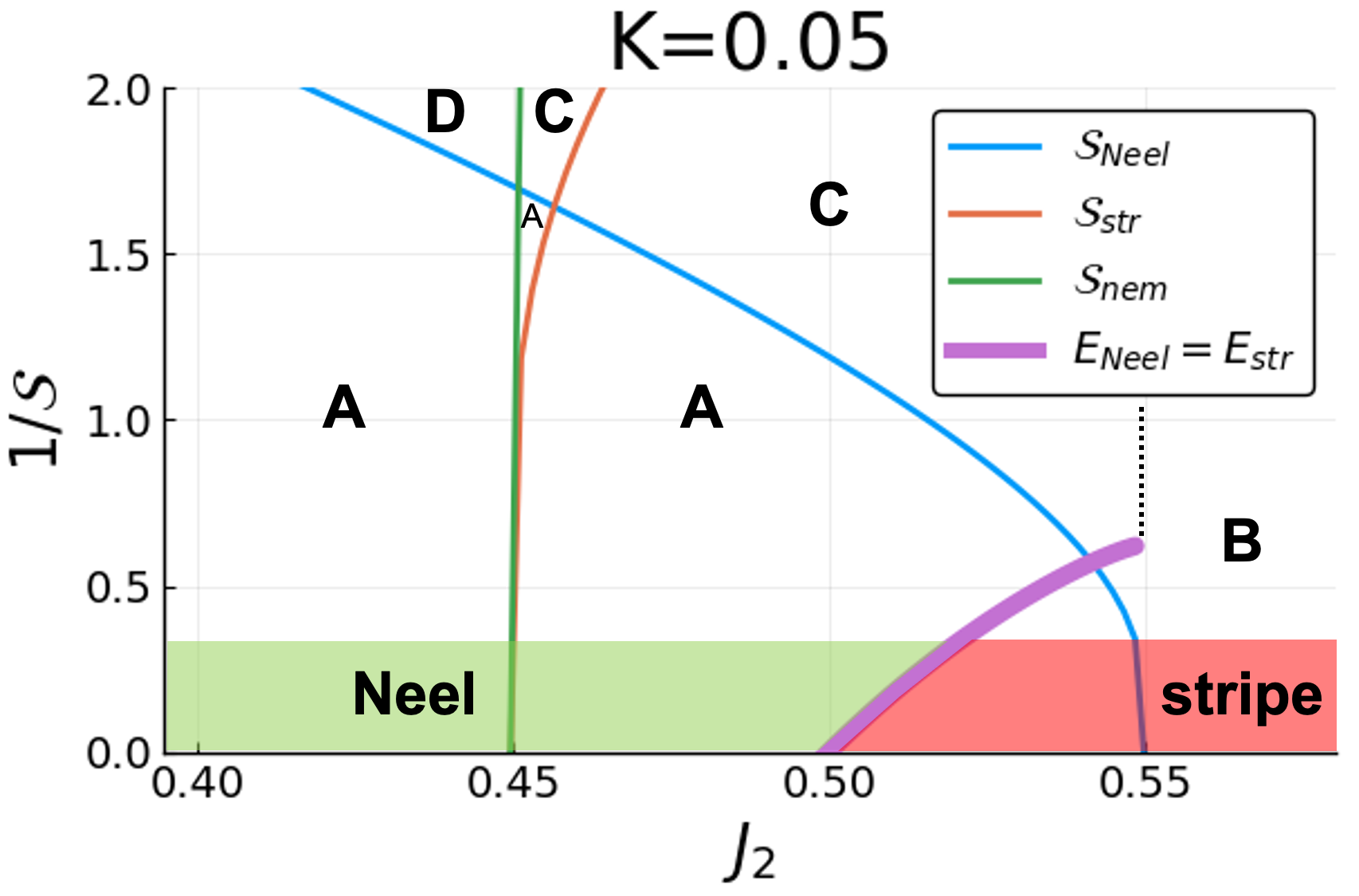}
\caption{%Phase diagram in the $J_2/J_1-1/S$ plane,
First order quantum corrections to various order parameters  as a function of $J_2/J_1$ for the square lattice with $K=0.05$. The thick solid purple line indicates a first order-boundary between the Neel and stripe phases, computed from Eq. \ref{firstorder}. The other solid lines represent $1/{\cal S}_a$ for $a=$ Neel (blue), stripe (orange) and nematic (green).  Interpreted as a generalized phase diagram, with $1/S$ along the y axis, the green and red regions in the large $S$ portion of the phase diagram represent the portion %of the inferred phase diagram 
 that can be determined without further argument. Other regions of the phase diagram are labeled with letters for use in the discussion in Sec. \ref{smallS}. 
}
\label{1spositiveK}
\end{figure}

Since the classical transition between the Neel and stripe phases is first order, the $1/S$ correction to the location of the phase boundary can be computed directly by identifying the point at $E_{Neel}=E_{str}$
\begin{equation}
E_{Neel}-E_{str} = -2(J_1-2J_2) +S^{-1}\left [{\cal E}_{Neel}-{\cal E}_{str}\right]+ \ldots
\label{firstorder}
\end{equation} %{\color{red} sign of ${\cal E}_{Neel}$ is changed to match Eq.2}
This is indicated by the heavy purple line in Fig. \ref{1spositiveK}. Note that quantum fluctuations stabilize the Neel state relative the stripe-ordered state. The line terminates at a value of $J_2$ at which the classical Neel state ceases to be metastable;  to determine the nature of the phase boundary at larger values of $1/S$ (smaller $S$) we will need to rely on indirect arguments, as we will discuss below. The remaining lines in Fig.  \ref{1spositiveK} show the calculated values of $1/{\cal S}_a$ vs. $J_2$.
As is clear from the figure, for all three orders, the values of $1/{\cal S}_a$ vanish as $J_2 \to $ a critical value that depends on the nature of the order involved.  This would seem to indicate that at this point the corresponding classical phase is unstable to quantum disordering even for arbitrary large $S$ so long as $S$ is not infinite.  However, for large $S$ (where our calculations are controlled) these putative quantum disordering transitions are pre-empted by the first order transition already discussed. In addition, it is worth noting that $\mathcal{S}_{nem}$ is negative for 
 intermediate $J_2$ ($J_2=0.5$ to $0.9$ at $K=0.05$), which means first order quantum correction could enhance the nematic order.
We will return to these results in Sec. \ref{smallS} below, where we will present arguments to determine the nature of the phases in various regions labeled by letters in Fig. \ref{1spositiveK}.

\begin{figure}[htb]
\includegraphics[width=8cm]{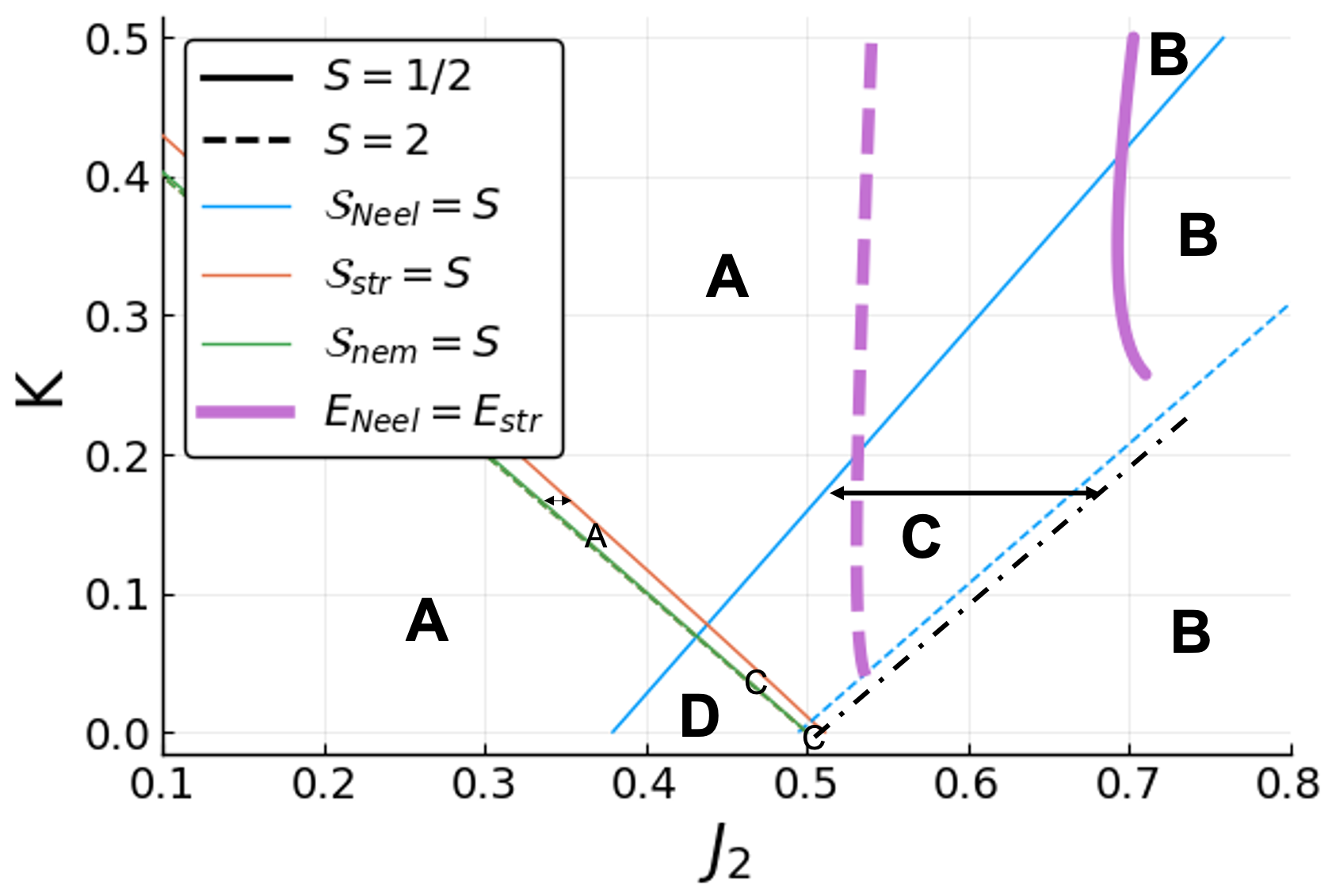}
\caption{
Contours (thin lines) of constant ${\cal S}_a(J_2,K)=S$ for the square lattice with $K>0$.  The solid lines are for $S=1/2$, and the dashed lines are for $S=2$. The heavy purple line indicates the contour along which $E_{Neel}=E_{str}$ as computed from Eq. \ref{firstorder}. 
The dashed-dotted line and letters identify different regions, as discussed in Sec. \ref{smallS}.
}
\label{f1}
\end{figure}

The K-dependence of the various quantities at fixed $S$ are shown in Fig.\ref{f1}. The thin lines indicate ${\cal S}_a =S$ and the heavy purple line marks the point at which one would conclude $E_{Neel}=E_{str}$ from the expressions computed to first order in $1/S$. The solid lines show results for $S=1/2$ while the dashed lines are for $S=2$. The dashed-dotted line is where the classical Neel state starts/ends to be metastable. It is not the true boundary between nematic and stripe regime, but will later be useful in Sec. \ref{smallS} to estimate the true boundary.

\subsubsection{Square lattice with $K<0$}
The relevant phases for $K<0$ are the Neel, spin-vortex crystal (SVC) and conical spin-vortex crystal (CSVC). 
All are states with non-zero sublattice magnetization, $m_a$, with $a=$Neel, $a=$SVC, and $a=$CSVC. 
In addition, we  define composite order parameters that capture specific aspects of the broken symmetry of various phases.  As with the more familiar case of nematic order arising as ``vestigial order\cite{nie}'' upon partial melting of a stripe-ordered state, it is possible to conceive\cite{fernandes} of phases with vestigial broken symmetries that arise by partial melting of, respectively, a SVC or a CSVC phase such that the primary order parameter vanishes but the composite order parameters remain finite. Specifically, we define  
\begin{equation}
\begin{split}
\vec {\bf O}_{SNVC}&\equiv \frac 1 {4NS^2}
\sum_{{\bf R}}e^{i{\bf Q}\cdot{\bf R}}\Big\langle {\vec S}_{{\bf R}+\hat{\bf y}}\times{\vec S}_{{\bf R}}+{\vec S}_{{\bf R}}\times{\vec S}_{{\bf R}+\hat{\bf x}}\\
&+{\vec S}_{{\bf R}+\hat{\bf x}}\times{\vec S}_{{\bf R}+\hat{\bf x}+\hat{\bf y} }+{\vec S}_{{\bf R}+\hat{\bf x}+\hat{\bf y}}\times{\vec S}_{{\bf R}+\hat{\bf y} }\Big\rangle\\
{\bf O}_{chir}&\equiv \frac 1 {\mathcal{N}_{chir}}
\sum_{{\bf R}}{\Big\langle{\vec S}_{\bf{R}}\cdot({\vec S}_{\bf{R+x}}\times{\vec S}_{\bf{R+y}})\Big\rangle}
\end{split}
\end{equation}
which we will refer to as the spin-nematic vortex crystal (SNVC) and chiral (chir) order parameters. %The (partial) sum for SNVC is among all set of spins $\vec{S}_i,i=1...4$ in closewise order within every other squares. The summation within the remaining half of squares can be shown to be exactly opposite. 
Here ${\bf Q} \equiv (\pi,\pi)$.
These order parameters are, respectively, an axial vector and a pseudo-scalar in spin-space, both normalized so that their magnitude is 1 in a corresponding classically ordered state. 
A phase with $\langle\vec  { \bf O}_{SNVC}\rangle\neq 0$ but $m_{SVC}=0$ thus breaks spin-rotational order in the same sense as a spin-nematic, as well as breaking translational symmetry.  (From a broken symmetry perspective, it is equivalent to a triplet d-density wave.\cite{tripletddw}.) A phase with $\langle{ \bf O}_{chir}\rangle\neq 0$ but $m_{CSVC}=0$ preserves spin-rotational and translational symmetries, but has net spin chirality. 

The dashed lines in Fig. \ref{classical} represent continuous transitions at $S=\infty$ between states  that break distinct symmetries; since the CSVC interpolates between the Neel and the SVC phase, such transitions are in principle consistent with Landau theory. However, what is not generic is that upon approaching the transition from either side, both phases cease  to be metastable ({\it i.e.} an appropriate spin-wave velocity vanishes) along the phase boundaries. Consequently, we expect that for finite  $1/S$, an intermediate region % in which 
with neither form of magnetic order occurs. 

\begin{figure}[h]
\centering
\includegraphics[width=8cm]{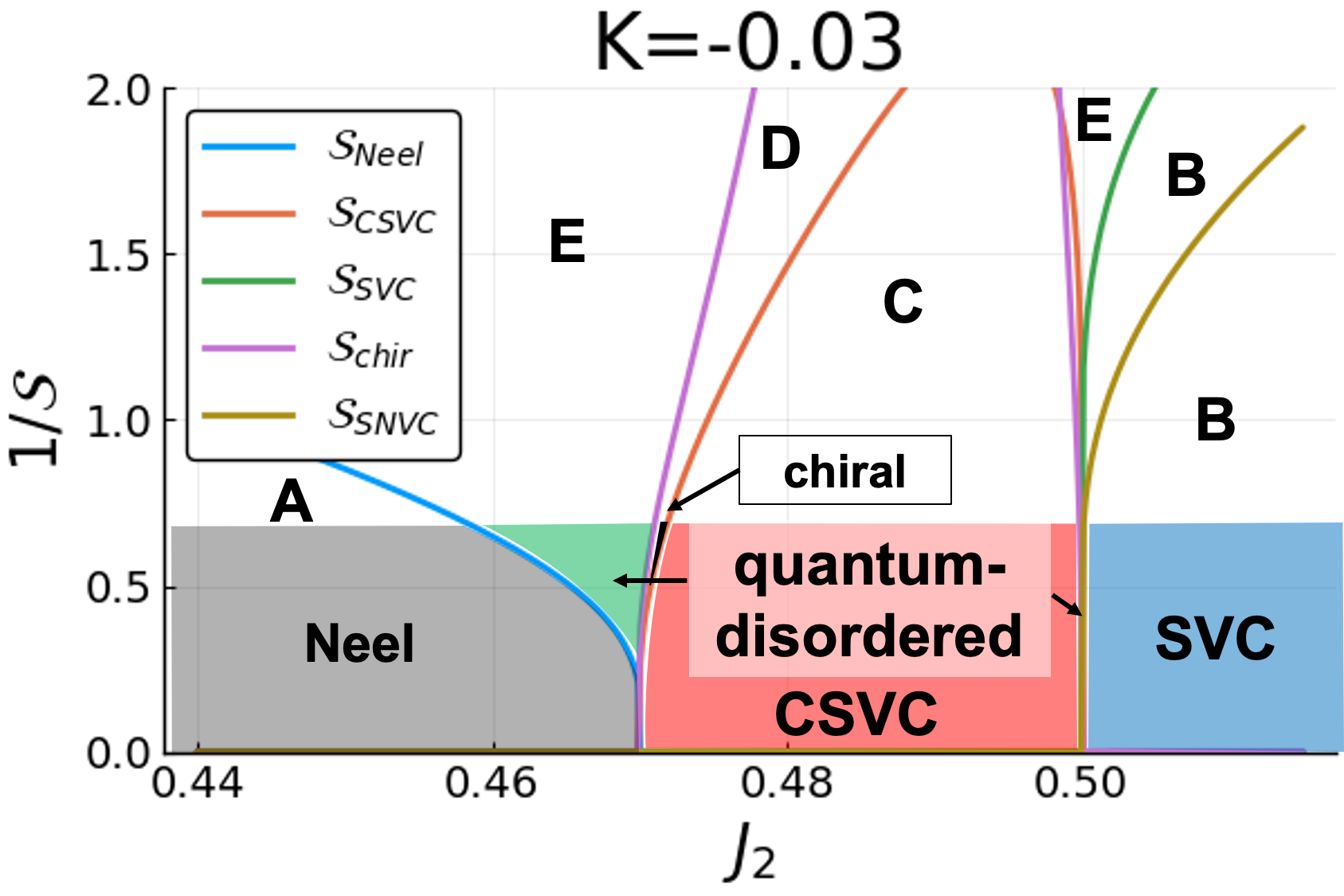}
\caption{%Phase diagram  in the $J_2/J_1-1/S$ plane,
First order quantum corrections to various order parameters  as a function of $J_2/J_1$ for square lattice with $K=-0.03$. The solid lines represent $1/{\cal S}_a$ for $a=$ Neel (blue), conical spin-vortex crystal (orange) and spin-vortex crystal (green), as well as for $a=$ %conical spin nematic vortex crystal
chiral (purple) and spin nematic vortex crystal (brown) phases. Interpreted as a generalized phase diagram, with $1/S$ along the y axis, the shaded regions in the large $S$ portion of the phase diagram represent the portion of the inferred phase diagram that can be determined without further argument. Other regions of the phase diagram are labeled with letters for use in later discussions of how to interpret the results for smaller $S$. }
\label{f3}
\end{figure}

Indeed, as shown in Fig. \ref{f3},  the curves of $1/{\cal S}_a$ vs $J_2$ for  the various phases diverge from each other slowly as $J_2$ is tuned away from its critical value, $J_{2c}(K)$ as %$a=$ Neel and CSVC order vanishes at the same position $J_2= J_1/2+K$ with contours of constant $1/{\cal S}_a$ which behave asymptotically as
\begin{equation}
\frac 1{{\cal S}_a} \sim \frac{ -G_a}{\ln\left| {J_2 - J_{2c}}\right| }\  {\rm as }\  \left| {J_2 - J_{2c}}\right| \to 0
\label{asymptote}
\end{equation}
where in this expression $J_{2c}= J_1/2+K$  near the convergence of the solid blue and solid orange lines (for $a=Neel$ and $a=CSVC$) and  $J_{2c}=J_1/2$ near the convergence of the solid orange and solid yellow  lines (for  $a=CSVC$ and $a=SVC$).  Note that for each value of $a$, this expression only applies as one approaches the critical value $J_{2c}$ from the appropriate direction and that in all cases $G_a>0$  with values that we compute explicitly in the appendix. 
% in the case of the solid blue and solid orange contours (for $a=Neel$ and $a=CSVC$) it applies to $a=CSVC$ for $J_2 > J_{2c}$ with $G_{CSVC} = ***$ and for $a=Neel$ for $J_2 < J_{2c}$ with with $G_{Neel} = ***$.  This leads, even at large $S$, to a narrow region (the green and dar blue shaded regions) between the solid blue and solid purple contours where one would expect that quantum fluctuations are sufficient to destroy all magnetic order.
These results strongly suggest that, even for large $S$, quantum disordered phases arise near $J_2 = J_1/2+K$ between the Neel and CSVC phases, shown as the green shaded regions in Fig. \ref{f3}, and near $J_2=J_1/2$ between the CSVC and SVC phases, which is too narrow to show up as a shaded region but is also indicated in the figure.
Note that the purple line marking the contour at which the extrapolated  chiral order vanishes lies inside a quantum disordered region  suggesting that there are (at least) two distinct phases here - one with  vestigial  chiral order and one that is more fully quantum disordered.    We will return to these results in Sec. \ref{smallS} below for smaller $S$, where we will present arguments to determine the nature of the phases in the various regions labelled by letters in Fig. \ref{f3}.

\begin{figure}[h]
\centering
\includegraphics[width=7cm]{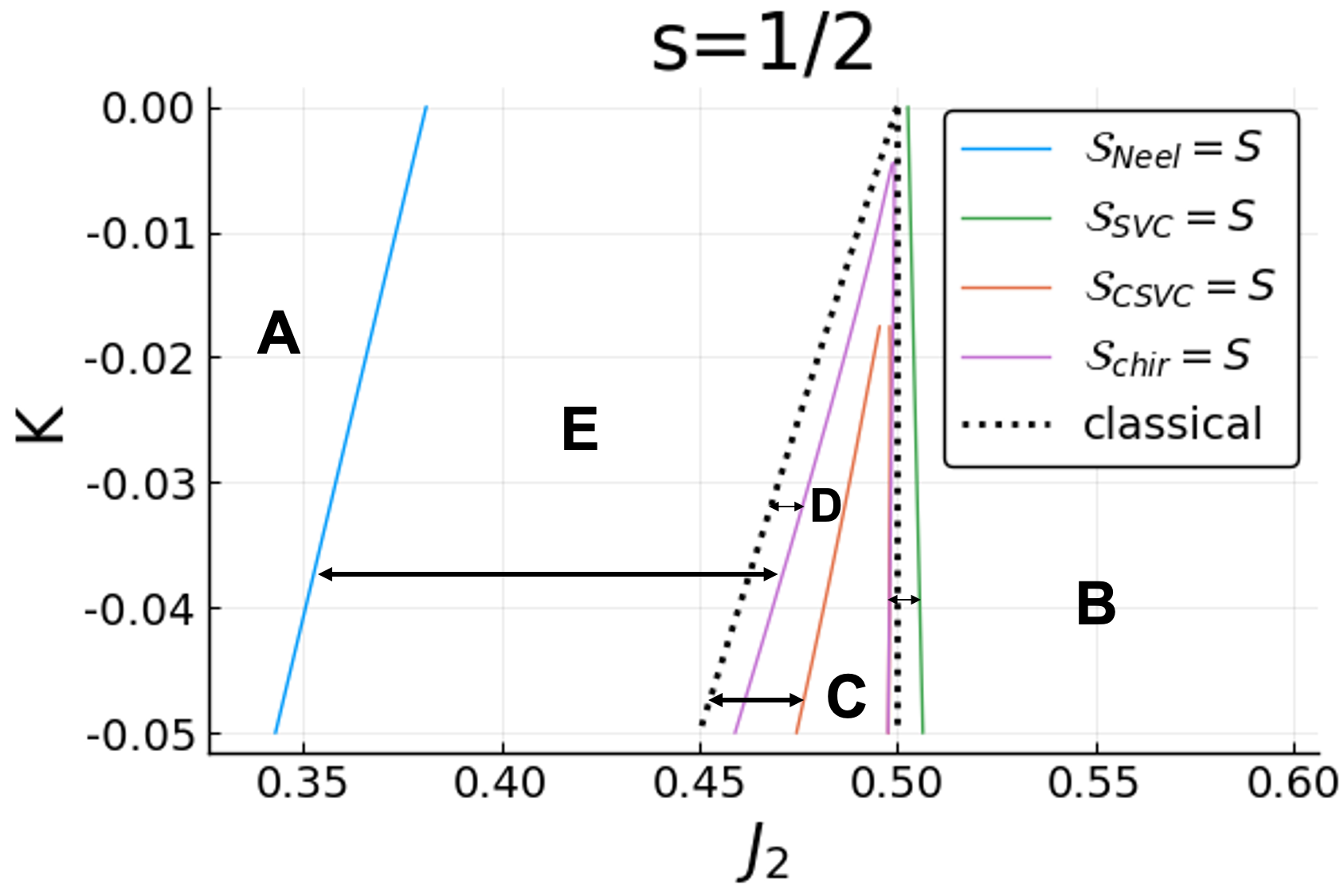}
\caption{Contours of constant ${\cal S}_a(J_2,K)%=S
=1/2$ for the square lattice with $K<0$. The letters identify different regions discussed in Sec. \ref{smallS}.}
\label{f4}
\end{figure}

The K-dependence of the various quantities at fixed spin $S=1/2$ are shown in Fig.\ref{f4}. Solid lines indicate ${\cal S}_a =S$ from the expressions computed to first order in $1/S$. Classical ($S\to \infty$) phase boundaries are added as dashed lines.

\begin{figure}[h]
\centering
\includegraphics[width=8cm]{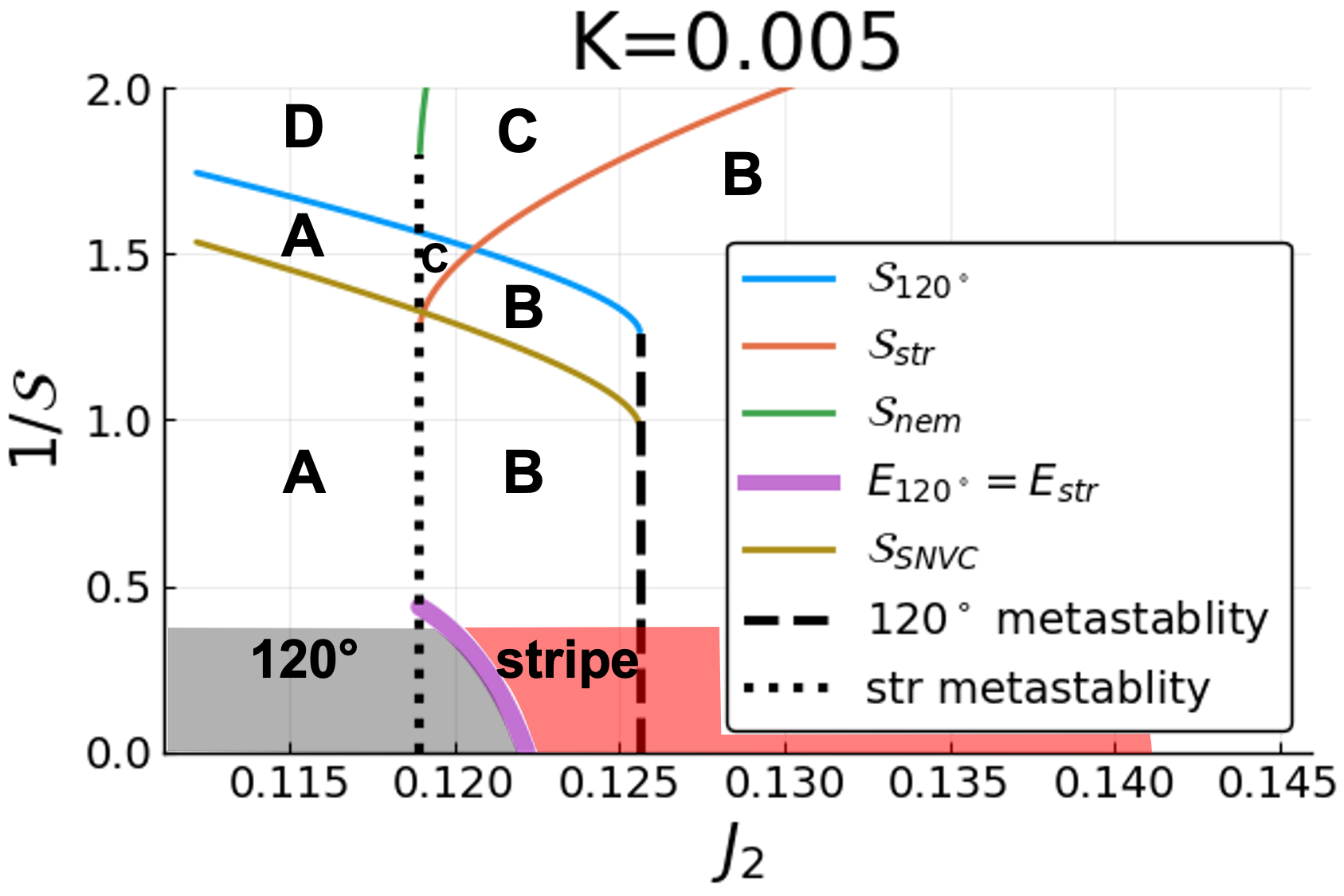}
\caption{%Phase diagram in $J_2/J_1-1/S$ plane, 
First order quantum corrections to various order parameters  as a function of $J_2/J_1$ for triangular lattice with $K=0.003$. The thick solid purple line indicates a first order-boundary between the $120^\circ$ and stripe phases, computed from Eq. \ref{Tfirstorder}. The other solid lines represent $1/{\cal S}_a$ for $a=$ $120^\circ$ (blue), stripe (orange), SNVC (brown) and nematic (green).  Interpreted as a generalized phase diagram, with $1/S$ along the y axis, the grey and red regions in the large $S$ portion of the phase diagram represent the portion of the inferred phase diagram that can be determined without further argument. Other regions of the phase diagram are labeled with letters for use in later discussions of how to interpret the results for smaller $S$.
%{\color{blue}  Why is it brown line here and a purple line in Fig. 5} 
 }
\label{Tf2}
\end{figure}

\subsubsection{Triangular lattice with $K>0$}

The two pertinent phases for $K>0$ are the $120^\circ$ and the stripe phase. Both are states with non-zero sublattice magnetization, $m_a$, with $a=120^\circ$ and $a=$str (for stripe order).
%
%Again, the 
The stripe phase also breaks the lattice 6-fold rotational symmetry in a manner that is characterized by the (three-state)  nematic order parameter,
\begin{eqnarray}
&&{\bf O}_{nem}\equiv \frac{1}{2NS^2}\sum_{\vec R} \Big\{\langle{S}_{\bf{R}}\cdot S_{\bf{R+\delta_1}}\rangle \\
&&+e^{i2\pi/3}\langle{S}_{\bf{R}}\cdot S_{\bf{R+\delta_2}}\rangle+e^{-i2\pi/3}\langle {S}_{\bf{R}}\cdot S_{\bf{R+\delta_3}}\rangle\Big\}%2S(S-\beta)
\ . \nonumber
\end{eqnarray}
Here, $\pm \bf{\delta}_i$ with $i=1$, 2, or 3 % the unit vector 
are the unit vectors on the triangular lattice, and ${\bf O}_{nem} =1$ in the classical stripe-ordered state in which the spins on site $\bf{R}+\bf{\delta}_2$ and $\bf{R}+\bf{\delta}_3$ have the %same orientation  in the stripe phase.
opposite spin orientation as the spins on sites ${\bf R}$ and ${\bf R}+{\bf \delta}_1$, while ${\bf O}_{nem}  =(1\pm i\sqrt{3})/2$ for the two other classical stripe ordered states.
%In much the same way as the 
Like the spin vortex crystal phase in the square lattice, $120^\circ$ phase also breaks the spin rotational symmetry. The vestigial phase can be characterized by the spin-nematic vortex crystal order:  % (a close relative of  triplet d-density wave order that has been considered in other contexts\cite{sudipddw}):
\begin{equation}
\begin{split}
\vec {\bf O}_{SNVC}&\equiv \frac 2 {3\sqrt{3}NS^2}\\
&\sum_{\triangleleft}\langle{\vec S}_1\times{\vec S}_2+{\vec S}_2\times{\vec S}_3+{\vec S}_3\times{\vec S}_1\rangle
\end{split}
\end{equation}
The sum is among all set of spins $(\vec{S}_1,\vec{S}_2,\vec{S}_3)$ in clockwise order (vorticity), within type A triangles ($\triangleleft$). The sum for spins also in clockwise order within type B triangles ($\triangleright$) can be shown to be exactly opposite. A phase with $\langle\vec  { \bf O}_{SNVC}\rangle\neq 0$ but $m_{120^\circ}=0$ thus breaks spin-rotational order, but, in contrast with the SNVC phase on the square lattice, it does not  break translational symmetry. 
%***I am up to here.***

Since the classical transition between the $120^\circ$ and stripe phase is first order, the $1/S$ correction to the location of the phase boundary can be computed directly by identifying the point at $E_{120^\circ}=E_{str}$
\begin{equation}
E_{120^\circ}-E_{str} = -0.5(J_1-8J_2) +S^{-1}\left [{\cal E}_{120^\circ}-{\cal E}_{str}\right]+ \ldots
\label{Tfirstorder}
\end{equation}
This is indicated by the heavy purple line in Fig.\ref{Tf2}.  Note that quantum fluctuations stabilize the stripe state relative the $120^\circ$ state. The line terminates at a value of $J_2$ at which the classical stripe state ceases to be metastable;  The remaining lines in Fig. \ref{Tf2} show the calculated values of $1/{\cal S}_a$ vs. $J_2$.
As is clear from the figure, for all four orders, $1/{\cal S}_a$ approaches non-zero values as $J_2 \to $ a critical value that depends on the nature of the order involved.

\begin{figure}[h]
\centering
\includegraphics[width=8cm]{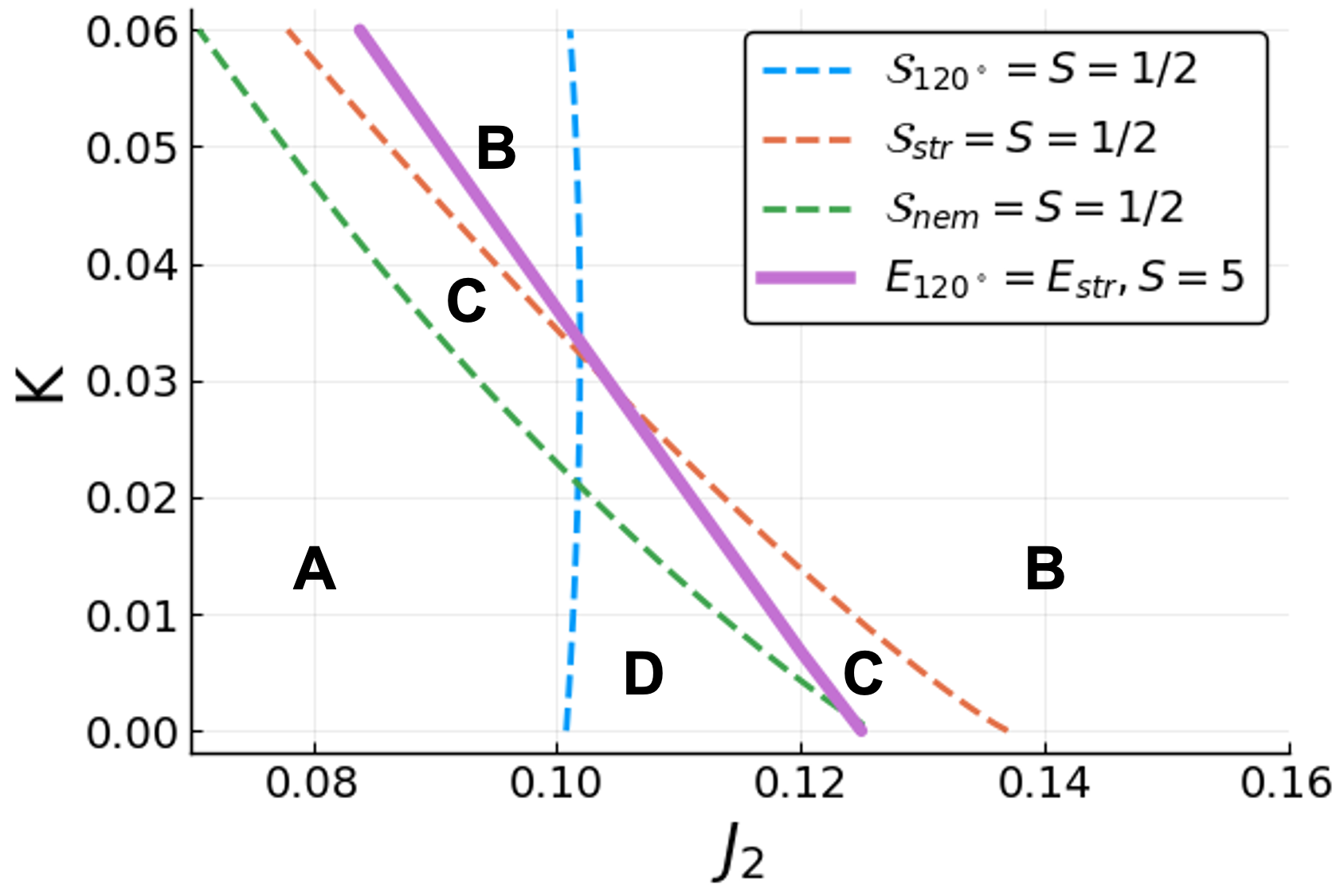}
\caption{Contours of constant ${\cal S}_a(J_2,K)=S$, as dashed lines for $S=1/2$. The solid heavy purple line indicates the contour along which $E_{120^\circ}=E_{str}$ as computed from Eq. \ref{Tfirstorder}, for $S=5$. The letters identify different regions, as discussed in Sec. \ref{smallS}.
}
\label{Ts1}
\end{figure}

The K-dependence of the various quantities at fixed %small spin $S=1/2$ and large spin $S=5$ is 
$S$ are shown in Fig.\ref{Ts1}.  Representative of the large spin case,
 the %solution of 
line in the $K-J_2$ plane along which $E_{120^\circ}=E_{str}$ for $S=5$ is marked by the heavy dashed purple line;  no solution of this equation exists in the small spin case, $S=1/2$.  %.  while ${\cal S}_a=S$ do not have solutions.} For small spin,
Conversely, % the solutions of 
the lines on which  ${\cal S}_a =S$ in the small spin case, $S=1/2$, are marked by the thin solid lines;  % but no such lines , while $E_{120^\circ}=E_{str}$ does not have solution.
no such lines exist  at large spin $S=5$.

\begin{figure}[h]
\centering
\includegraphics[width=8cm]{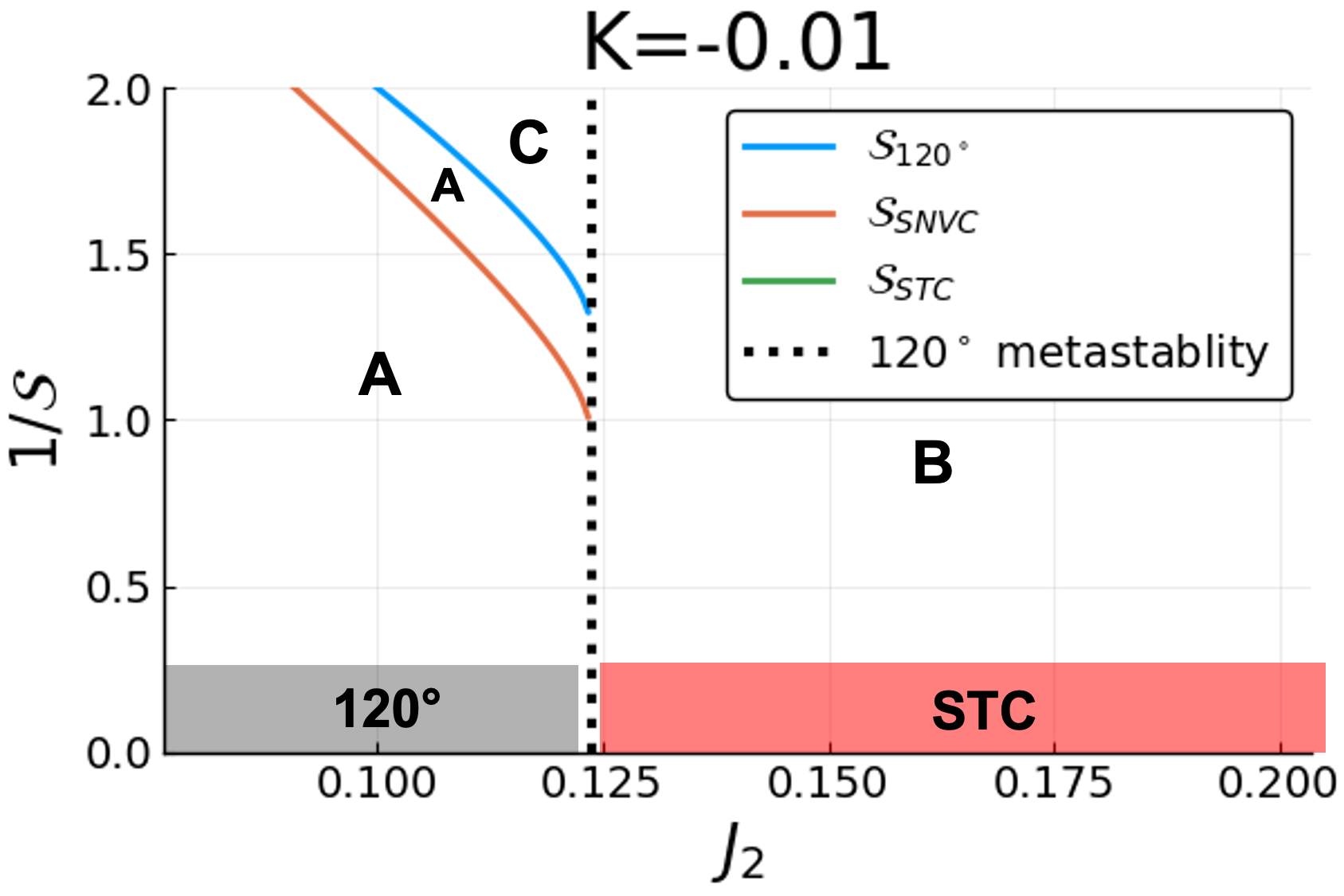}
\caption{Phase diagram in $J_2/J_1-1/S$ plane, for $K=-0.01$. The solid lines represent $1/{\cal S}_a$ for $a=$ $120^\circ$(blue), SNVC(orange) and STC phase (above $S=1/2$ boundary).
The shaded regions in the large $S$ portion of the phase diagram represent the portion of the inferred phase diagram that can be determined without further argument. Other regions of the phase diagram are labeled with letters for use in later discussions of how to interpret the results for smaller $S$.
%{\color{blue}  Again, I the dashed purple line should be a thick solid line}
}
\label{T1snK}
\end{figure}

\subsubsection{Triangular lattice with $K<0$}

The two pertinent phases for $K<0$ are the $120^\circ$ and the spin-tetrahedron
crystal phase. Both are states with non-zero sublattice magnetization, $m_a$, with  $a=120^\circ$ and $a=$STC (for spin-tetrahedron crystal order).

Since the classical transition between the $120^\circ$ and STC phase is first order, the $1/S$ correction to the location of the phase boundary can be computed directly by identifying the point at which $E_{120^\circ}=E_{STC}$.
In the classical $S\to \infty$ limit, this occurs at the limit of metastability of the $120^\circ$ phase.  Moreover, in the entire regime in which % It turns out that, when 
 both states are classically metastable, the $120^\circ$ phase always has a lower energy %under
 even when first order quantum corrections are included.  Thus, this first order phase boundary does not vary with $S$ to first order in $1/S$;  %. Therefore, at large $S$, there will be a first order phase transition
 the first order boundary between $120^\circ$ phase and STC phases occurs at $J_2=\frac{J_1}{8}+\frac{K}{8}$ in Fig.\ref{T1snK}. % where classical $120^\circ$ state ceases to be metastable.

\begin{figure}[h]
\centering
\includegraphics[width=8cm]{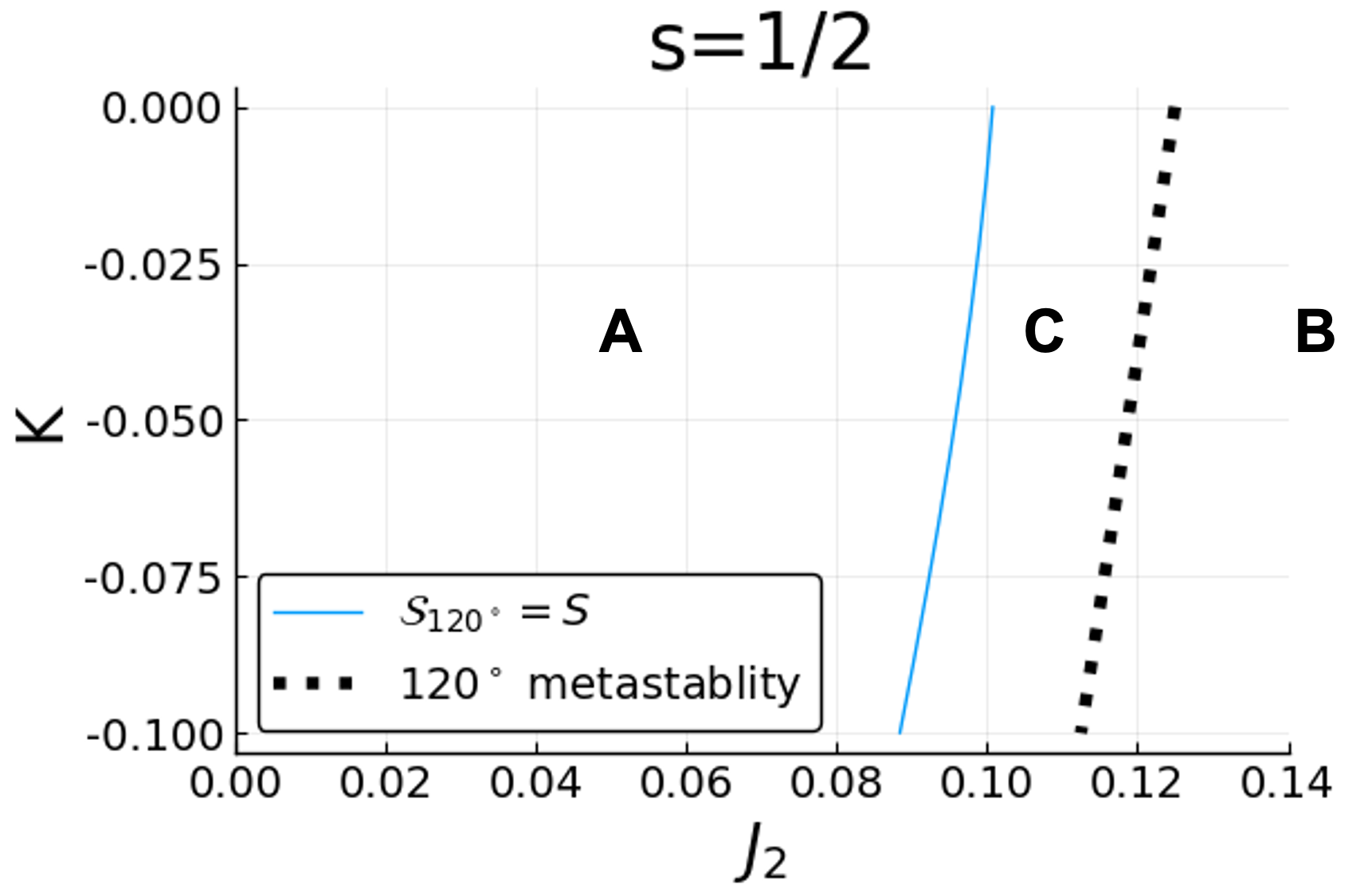}
\caption{Contours of constant ${\cal S}_a(J_2,K)=S$, for $S=1/2$. 
The letters identify different regions discussed in Sec. \ref{smallS}.
Black line is where $120^\circ$ phase ceases to be metastable.}
\label{Ts5kn}
\end{figure}

The K-dependence of the various quantities at fixed spin $S=1/2$ are shown in Fig.\ref{Ts5kn}. Thin solid lines indicate ${\cal S}_a =S$ from the expressions computed to first order in $1/S$. 

\section{Extrapolation to smaller $S$}
\label{smallS}
While there are surely dangers involved, it is worth-while extrapolating the results that are controlled at large $S$ to form at least a conjectural completion of the phase diagrams, as shown in Fig. \ref{schematic}.  This  also allows us to make contact with a host of numerical studies that have been performed on the same models for the cases of $S=1/2$ and $S=1$.  In this section, we outline the logic that leads to this figure.  In all cases, when we refer to ``the ground state energy'' or ``the magnitude of the order parameter'' we are implicitly referred to quantities that are computed to leading order in $1/S$.  For example, when we refer to a region of the phase diagram in which the Neel order parameter vanishes, we mean a region where $S < {\cal S}_{Neel}$, i.e. where the extrapolated magnitude of the order parameter would be negative.

\subsection{Square lattice with $K>0$}
%Now we extrapolate the result to smaller $S$ for 
Our analysis of the % $K>0$ on
 square lattice with $K>0$ is based on identifying the nature of the ground-state in the various % Regions
regions shown in Fig. \ref{1spositiveK} which are %now 
labelled by different letters. %For
Everywhere to the left of the solid purple line (i.e. regions A, C, and D) 
%In region (A) below the blue line $1/{\cal S}_{Neel}$, %our extrapolation says
%taking at face value the extrapolated values of the quantum fluctuations leads to the conclusion that the Neel order parameter is non-zero. Comparing 
the energy of the Neel state is lower than the stripe phase, implying that the stripe phase is excluded in all these regions.  
%energies evaluated to first order in $1/S$, one also finds that the Neel state is %also energetically preferred over the stripe state/ 
In region A (i.e. below the blue line) the  magnitude of $m_{Neel}$ is positive, so we identify this as being approximately the region in which Neel order survives quantum fluctuations.
 %Thus we %believe conjecture that region (A) is in tne Neel phase. 
 Conversely, in regions C and D, $m_{Neel}$ vanishes, which we interpret as meaning that no magnetic order survives in either region.  Indeed, in region D all the orders we have considered are precluded, so we identify it as a fully quantum disordered regime.  On the other hand, in region C quantum fluctuations of the nematic order do not vanish cause it to vanish, so we conjecture that this corresponds to a nematic phase.
Similarly all of region (B) is likely stripe ordered, since the stripe state is energetically preferred, and its order parameter including first order quantum corrections is non-zero. 
%Region (C) is not in the Neel nor stripe phase, since the extrapolated value of the Neel order parameter vanishes %from extrapolation, 
%in this region and the stripe state has a higher energy than the Neel state. %The
%However, even including first order quantum fluctuations, the nematic order parameter here is non-zero. %Even 
%Although we do not %know
%have a direct way of computing the energy of the nematic state, %we think 
%these results are suggestive that Region (C) %may at least have similar topological feature as the
%likely exhibits vistigial nematic phase.
%In region (D), the values of all of the above order parameters vanishes, so %it is in the
%we identify this regions as a  quantum disordered phase.

%{\color{blue} I am here.}

The nature of phase transitions can now be %estimated.
considered. %For example, t
The %phase
 transition indicated by the heavy purple line in Fig. \ref{1spositiveK} between the Neel and stripe phases is %a first order transition
 first order, and is unambiguously calculable when $S$ is large. %which is from large $S$ calculation that does not require extrapolation. 
 Since we have identified the  phase transitions between different regions separated by thin solid lines as the points at which quantum fluctuations become large enough that one or another order parameter vanishes, we have implicitly assumed that these transitions are all continuous.
% The phase transitions between any ordered phase and  a quantum disordered phase %should be continuous, 
% is probably continuous, klsince the order parameters $|{\bf O}_a|$ continuously vanishes as approaching those phase boundaries. 
For the cases of the Neel to disordered (A to D) and  the nematic to disordered (C to D), these correspond to reasonable Landau-allowed order to disorder transitions.  However, the implied Neel to nematic transition (A to C) is not Landau allowed, and indeed to the extent that the first order in $1/S$ expressions can be trusted, the nematic order would have a finite jump across this transition.  While, as discussed in Sec. \ref{subtleties}, under special conditions, a ``beyond Landau'' continuous deconfined quantum critical transition\cite{deconfined} between a Neel and a quantum nematic paramagnet is possible\cite{Fawang}, far more likely is that  the transition between these two phases is first order  and probably not quite at the same point as the solid blue line.
 %Phase boundary between nematic and Neel phase should be first order phase transition, since the broken symmetries in these two phases are not subgroup of one and another. 
The nature of the phase transition between the nematic and stripe phases is also unclear.  This is Landau-allowed to be continuous although at least where it occurs along the heavy purple line it is probably first order. If we were to extend the range of parameters shown in Fig. \ref{1spositiveK} we would find that the orange line $1/{\cal S}_{str}$ and the green line $1/{\cal S}_{nem}$ cross at $J_2\approx{3J_1}$;  we interpret this crossing as a bicritical  point marking    the end of the first order stripe to nematic phase boundary, beyond which a direct stripe to quantum disordered transition is expected.
 
 %{\color{red} We may need to discuss a little bit about the use of first order transitions for boundary between nematic and stripe phase in Fig1.}

The K-dependence of phase diagram can %now 
be obtained %in the same extrapolation. 
from the same sort of analysis.  Contours of $\mathcal{S}_a=S$ and lines of $E_{a}=E_{b}$ in Fig. \ref{f1} are now used to estimate the %true 
 phase boundaries. Phases (A) Neel, (B) stripe, (C) nematic %-like 
  and (D) quantum disordered are explicitly labelled for $S=1/2$. %For the same reason mentioned above, phase boundary between stripe and nematic phase should be on the right of the current one. 
 For small $K$, the width of the "nematic-like'' phase (C), and of the quantum disordered phase (D) -- indicated by the black arrows in the main panel of Fig.\ref{f1}), become exponentially small as $S$ increases, and so are invisible for $S=2$.

 Together, these considerations lead to the schematic phase diagram %is then 
shown in Fig.\ref{1a}. 

%{\color{blue}  I am up to here.}

\subsection{Square lattice with $K<0$}

%We now identify the nature of the ground-state on 
For the square lattice with $K<0$, %in the various regions shown in 
the analysis that leads from Figs. \ref{f3} and \ref{f4}  to the schematic phase diagram in  Fig.\ref{1b} is relatively straightforward.  %which are labelled by different letters. It turns out that the analysis is mathematically the same as the discussion for large $S$. 
Specifically, as can be seen in
%This can be verified in 
 Fig. \ref{f3}, %where all the regions at large $S$ continuously extends to smaller $S$. We thus do not repeat the discussion here.
 the basic topology of the phase diagram is already established when $S$ is large, and hence can be interpreted without need of extrapolating to smaller $S$.  Thus % The 
 regions A, B, C, D, and E  correspond to Neel, SVC, CSVC, chiral and quantum disordered phases, respectively.

The nature of phase transitions can now be considered.  %
Again, since we have identified the phase transitions %between different regions separated by thin solid lines 
as the points at which quantum fluctuations become large enough that one or another order parameter vanishes, we have implicitly assumed that these transitions are all continuous. All the phase transitions here are Landau-allowed to be continuous.
The K-dependence phase diagram can be obtained from the same sort of analysis. %Contours $\mathcal{S}_a=S$ in Fig.\ref{f4} are now used to estimate the phase boundaries. Phases (A) Neel, (B) SVC, (C) CSVC, (D) chiral and (E) quantum disordered phase are explicitly labeled for $S=1/2$. Classical phase boundaries are added as black dashed line for comparison. Similar to $K>0$ system,
In common with the $K>0$ case, the width of the quantum disordered phase (E), and of the chiral phase(D) -- indicated by black arrow in Fig.\ref{f4} -- decreases exponentially as S increases. However, one noticeable difference is that the quantum disordered regime is present even at large $|K|$ in the phase diagram with $K<0$. 
Note that both CSVC and chiral phases requires sufficiently large $|K|$ to develop non-zero %value for their  
 order parameters. Therefore, at small $|K|$, the width of quantum disordered phase %will
  increase as $|K|$ increases.  %, which is different from the case for $K>0$. This difference
  This can be traced back to the fact that ``order by disorder" phenomena\cite{chandracolemanlarkin} %cooperate with attractive biquadratic interaction ($K>0$) 
  effectively add\cite{henley} a positive contribution to $K$  %to further 
  which tends to stabilize the two collinear states (Neel and stripe). %While for $K<0$, order by disorder phenomenon works against the repulsive biquadratic and disfavors the non-collinear states (SVC and CSVC).

These consideration lead to the  schematic phase diagram shown in Fig.\ref{1b}, in which phase boundary between chiral and CSVC phase have been extended and connected to larger $1/S$  for artistic reasons.

%{\color{blue} \subsection{Square lattice with $K=0$}
%**  I think this subsection can be omitted - it is implicitly contained in the above discussion.**

%The remaining phase boundary for square lattice is between stripe phase and SVC phase. Classically, this phase transition is first order at $K=0,\,J_2>J_1/2$. The classical stripe phase is metastable for $K\geq{0}$, while the classical SVC state is metastable for $K\leq{0}$. Thus, we cannot directly compare the ground state energies at any non-zero $K$. However, one could estimate the shift of the phase transition line, by comparing their ground state energies at $K=0$. For example, at $J_2=J_1$ and $K=0$, the difference in energy density is $E_{SVC}-E_{str}=0.03/S$. Therefore, the phase transition line from stripe phase to SVC phase is shifted towards SVC phase, by an amount proportional to $1/S$.}

\subsection{Triangular lattice with $K>0$}
%Our analysis of the triangular lattice with $K>0$ is based on identifying the nature of the ground-state in the various regions shown in
The schematic phase diagram for the triangular lattice with $K>0$ shown in in Fig. \ref{1c} is obtained by identifying the most likely phase corresponding to the different labeled retions in  Figs. \ref{Tf2} and \ref{Ts1} which are labelled by different letters. Everywhere to the left of the solid purple line in Fig. \ref{Tf2}  (i.e. regions A and D) the energy (%if exists
where it can be computed) of the $120^\circ$ state is lower than that of the stripe phase, implying that the stripe phase is excluded. % in all these regions.  
In region A (i.e. below the blue line) the  magnitude of $m_{120^\circ}$ is positive, so we identify this as the region in which $120^\circ$ order survives quantum fluctuations.
However, in regions D, $m_{120^\circ}$ vanishes, which we interpret as meaning that no magnetic order survives in %either 
this region.  Indeed, in region D all the orders we have considered are precluded, so we identify it as a fully quantum disordered regime.  
Similarly all of region B is likely stripe ordered, since the stripe state is energetically preferred, and its order parameter including first order quantum corrections is non-zero. In region C, the nematic order does not vanish, while stripe order vanishes and $m_{120^\circ}$ phase is energetically excluded; so we conjecture that this corresponds to a nematic phase.

The nature of phase transitions can now be considered. The analysis closely parallels that of the square lattice with $K>0$.  The transition indicated by the heavy purple line in Fig. \ref{Tf2} between the $m_{120^\circ}$ and stripe phases is first order, and is unambiguously calculable when $S$ is large. Since we have identified the  phase transitions between different regions separated by thin solid lines as the points at which quantum fluctuations become large enough that one or another order parameter vanishes, we have implicitly assumed that these transitions are all continuous.
For the cases of the $m_{120^\circ}$ to disordered (A to D) and  the nematic to disordered (C to D), these correspond to reasonable Landau-allowed order to disorder transitions.  However, the implied $m_{120^\circ}$ to nematic transition (A to C) is not Landau allowed, and indeed to the extent that the first order in $1/S$ expressions can be trusted, the nematic order would have a finite jump across this transition.  %While, as discussed in Sec. \ref{subtleties}, under special conditions, a ``beyond Landau'' continuous deconfined quantum critical transition\cite{deconfined} between a Neel and a quantum nematic paramagnet is possible\cite{Fawang}, far more
It is thus  likely  that  the transition between these two phases is first order  and probably not quite at the same point as the solid blue line.
The nature of the phase transition between nematic and stripe phase is also unclear.  This is Landau-allowed to be continuous. Different from square lattice, we are unable to extend the range of parameters shown in Fig. \ref{Tf2} to larger $J_2$, since there are other relevant phase.  Thus, we are unable to confirm the existence of a bicritical point marking the end of the first order stripe to nematic phase boundary, beyond which a direct stripe to quantum disordered transition is expected.

The K-dependence of phase diagram can be  obtained from the same sort of analysis.  Contours of $\mathcal{S}_a=S$ for $S=1/2$ (dashed lines) and line of $E_{a}=E_{b}$ for $S=5$ (solid line) in Fig. \ref{Ts1} are now used to estimate the phase boundaries. Phases (A) $120^\circ$, (B) stripe, (C)nematic-like and (D)quantum disordered are explicitly labelled for $S=1/2$. Contrary to the results of square lattice, since quantum correction $\mathcal{S}$ to order parameter does not have the logarithmic divergence as $J_2$ approaches the critical value $J_1/8$, the width of the nematic and quantum disordered phase does not decay exponentially as S increases. It has been shown in Fig.\ref{Tf2} that these two phases do not appear at large spin.

\subsection{Triangular lattice with $K<0$}
Finally, we use the results in Figs. \ref{T1snK} and Fig.\ref{Ts5kn} to construct the qualitative phase diagram for the triangular lattice with $K<0$ shown in Fig. \ref{1d}.
%Now we extrapolate the result to smaller $S$ for $K>0$ on triangular lattice. The results are summarized in Fig. \ref{T1snK}, where regions are labeled by letters. 
 It should be noted that the solid line  in Fig.  \ref{T1snK}  indicating  the value of $1/{\cal S}_a$ for $a=$STC  %phase is above $S=1/2$ boundary. 
is not shown since $1/{\cal S}_{STC}> 2$ corresponding to an unphysical value of $S < 1/2$.   For region A under the blue line %$1/{\cal S}_{120^\circ}$, our extrapolation says that
  the $120^\circ$ order parameter is apparently non-zero. %Since $120^\circ$ state , if metastable, is always energetically preferred over STC state, we think region (A) is in 
  We thus identify region A with the $120^\circ$ phase. Region B is identified with the STC phase, since only the STC state is metastable, and its order parameter is estimated to be  non-zero.
Region C is not in the $120^\circ$ nor the STC phase, since the $120^\circ$ order parameter vanishes, % from extrapolation, 
 and  the STC state has a higher energy than the $120^\circ$ state. It may be in a quantum disordered phase, or another symmetry breaking vestigial phases that we have not considered.  The K-dependence phase diagram at fixed $S=1/2$ is summarized in Fig.\ref{Ts5kn}. As is the case with the square lattice, quantum fluctuations tend to stabilize the positive $K$ phases relative to those with negative $K$ as a form of order from disorder.

The phase boundaries are now given by two lines. One is the blue line for $1/\mathcal{S}_{120^\circ}$ in Fig.\ref{T1snK}, that separates the $120^\circ$ phase and the quantum disordered phase. Another one is at $J_2=\frac{J_1}{8}+\frac{K}{8}$, where classical $120^\circ$ state ceases to be metastable, that separate STC phase with other two phases. 
%
%The nature of above phase transition can now be estimated. 
The phase transition between the $120^\circ$ phase and the quantum disordered phase should be continuous, since the order parameter ${\bf O}_{120^\circ}$ vanishes continuously upon approaching the phase boundary. The phase transitions between STC phase and the other two phases ($120^\circ$ and quantum disordered phase) should be first-order, since the STC order parameter ${\bf O}_{STC}$ is non-zero upon approaching the phase boundaries.  

%The K-dependence phase diagram at fixed $S=1/2$ is summarized in Fig.\ref{Ts5kn}. Phases (A) $120^\circ$, (B) STC and (C)quantum disordered phase are explicitly labeled. 

%{\color{blue} \subsection{Triangular lattice with $K=0$}
%***omit this subsection

%The remaining phase boundary for triangular lattice is between stripe phase and STC phase. Classically, this phase transition is first order at $K=0$. Similar as in the square lattice, we could estimate the shift of phase boundary by computing first order quantum correction to their ground state energy at K = 0. The resulted phase transition line is shifted towards the STC phase by an amount of O(1/S).}

\subsection{Comments on the validity of the extrapolation}
Based on the above extrapolation of %our results on
the  first order in $1/S$ expansion to smaller $S$, we obtained the schematic phase diagrams in Fig.\ref{schematic}. As stated in the beginning of this section, the extrapolation is dangerous and its validity cannot be guaranteed. However, in this subsection, we would like to comment on the validity of this extrapolation, by comparing the results fn the first order $1/S$ expansion with the second order the $1/S$ expansion.
Since the $1/S$ expansion is an asymptotic series,\cite{sudip} %even if the higher order results are extremely close to the first order results,  it is still not guaranteed to converge at all orders. 
where the first and second order results differ substantially, it is far from clear which is closer to the correct answer.  Therefore, this comparison only functions as  a ``comment" rather than any systematic proof of the validity of the above extrapolations.

For illustration purpose, we only focus on the stability of the stripe phase on the square lattice with $K>0$. This has been calculated in \cite{Majumdar2010}, for the case without a biquadratic interaction. In general, the biquadratic interaction can be approximated by the following quadratic interaction. 
\begin{equation}
(S_i\cdot{S_j})^2\approx\left[2(S_i\cdot{S_j})\langle{S}_i\cdot{S_j}\rangle-\langle{S}_i\cdot{S_j}\rangle^2\right] \ ,
\end{equation}
 and if the classical configuration is collinear, $\langle{S}_i\cdot{S_j}\rangle$ can then be replaced by its classical expectation value. For the stripe phase, the above mean-field approximation leads an anisotropic term, which has already been included in the previous study \cite{Majumdar2010}.

\begin{figure}[htb]
\includegraphics[width=6cm]{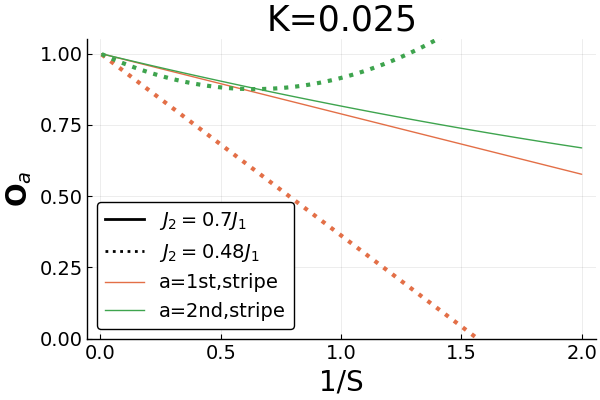}
\caption{Stripe order parameter ${\bf O}_{str}$, calculated from 1st and 2nd quantum correction as a function of $1/S$. Solid lines are for $J_2=0.7J_1$, and dashed lines are for $J_2=0.48J_1$, under the same $K$. 
%{ \color{blue}  I suggest dropping the classical lines.  To zeroth order in $1/S$ all of these quantities are equal to 1 by definition so this conveys no information.  Also, I wonder if it is possible to show data in which the extrapolation is more complicated, i.e. with instead of J2=1/2 have J2 0.455 or something close enough to the critical value of J2 that the first order curve extrapolates to zero before S reaches 1/2.}
}
\label{f2}
\end{figure}

As a function of $1/S$,  the stripe order parameter ${\bf O}_{str}$ under zeroth order (classical result, blue line), first order (orange line) and second order $1/S$ (green line), as well as nematic order parameter ${\bf O}_{nem}$ under first order (purple line) $1/S$ expansion are plotted in Fig.\ref{f2}.
Solid lines are for $J_2=0.7J_1$, and dashed lines are for $J_2=0.5J_1$. 
As $1/S$ grows, the second order results starts deviating from the first order quantum correction. The first order result on $\bf{O}_a$ should not be trusted if the deviation is big. Therefore, as $J_2\to{J}_1/2-K$, the order parameter $\bf{O}_a$ from first order quantum correction becomes trustable only at large spin. 

In our previous extrapolations, we use $\mathcal{S}$ to estimate the true phase boundary $\mathcal{S}_{crit}$. This estimation in general works better at larger spin and further away from the critical point.

\section{%Additional Subtleties
Beyond spin-wave analysis}
\label{subtleties}

There are a large number of additional subtleties that we have overlooked in the present analysis.  We have treated $1/S$ as a continuous parameter that tunes the extent of the quantum oscillations -- this is very similar in spirit to the classic approach of Ref. \onlinecite{chn}.  However,$S$ is in fact a discrete variable, and  there can be differences in the physics depending on whether it is integer or half-integer,\cite{haldane,readsachdev,senthilfisher} and even whether it is an even integer or an odd integer\cite{aklt,Fawang}.  This can effect the nature of the allowed phases and opens up the possibility of exotic, beyond Landau deconfined quantum phase transitions.\cite{deconfined}  For instance, generalizations of the famous Lieb-Schulz-Mattis theorem\cite{lsm,affleck,hastings}, imply that the disordered phase for a half-integer spin must either have a broken symmetry (e.g. exhibit valence-bond-crystalline order) or be one or another of quantum spin liquid with   topological order. 

Since most of the numerical studies to date have been carried out for $S=1/2$, or, to a lesser extent, for $S=1$, these additional subtleties are likely to be significant.  However, we can still distinguish magnetically ordered phases from quantum disordered phases.  Moreover, within the regime of quantum disordered phases, we may be able to distinguish those that exhibit broken symmetries as a form of vestigial order, if the order is accompanied by  reasonably long but still finite range correlations that reflect the structure of a nearby magnetically order, vs. broken symmetries (such as the before mentioned topological order) that are more readily identified with topological terms in the effective field theory\cite{senthilfisher}, rather than with any proximate magnetically ordered state.

%Vast amount of efforts have been put on
Intense effort and enormous creativity has been marshaled for the numerical search for intermediate non-magnetic quantum-disordered states in the $J_1-J_2$ models (wiht $K=0$). 
For $S=1/2$ on the square \cite{richter2015spin,gong2014plaquette,capriotti2001resonating,iqbal2016intertwined,hu2013direct,li2012gapped,morita2015quantum,wang2016tensor,jiang2012spin,figueirido1990exact} and triangular lattice \cite{iqbal2016spin,hu2015competing,kaneko2014gapless,zhu2015spin}, various of numerical works have confirmed the existence of an intermediate quantum disordered state(s). However, the nature of the intermediate state(s) is still under debate. For $S=1$ on a square lattice, there is contradictory evidence  concerning the existence of  intermediate phases; \cite{haghshenas2018quantum,Bishop_2008,jiang2009phase} %{\color{red} (while l did not find numerical works on $J_1-J_2$ model with $S=1$ on triangular lattice, double check needed?).} 
notably, in the study\cite{jiang2009phase} that is most strongly indicative of the occurrence of such a phase, it is found to occur in an exceedingly narrow range of $J_2$ and to have a clear nematic character.  We are unaware of any studies of the $S=1$ model on the triangular lattice.
Since finite S effectively adds a positive K, one should compare the above numerical studies with the phase diagrams in our work with $K>0$. At least in terms of the general topology of the phase diagram, the results of the existing numerical studies appear to be consistent with those shown in Figs. \ref{schematic} a and c. An interesting future direction for numerical studies suggested by the present study is to investigate the model with negative $K$ (or $K'$ for $S=1/2$) where % could be a decent numerical platform to study the known intermediate phases, and potentially search for new quantum disordered phases.
the quantum disordered phases are found to persist to large $S$ and to have a broader region of stability.

\section{conclusion}
We analyzed Heisenberg model on square and triangular lattice with nearest neighbor and next-nearest neighbor quadratic interactions, as well as nearest-neighbor biquadratic interaction at zero temperature. We analyzed the effect of the biquadratic interaction, and obtained phase diagram under first order quantum correction using linear spin wave theory. We compared our results on repulsive biquadratic interaction with the previously works on attractive interaction.

We found that the classical first order transition under attractive biquadratic interaction is preserved at large spin. The classical continuous phase transitions under repulsive biquadratic interaction is however replaced by quantum disordered regions. Those quantum disorder regions exist for arbitrary $S<\infty$ and reasonably small K.
We computed first order quantum corrections to short-range order parameters for non-collinear and non-coplanar states under repulsive biquadratic interaction, and observed the short-range ordered conical vortex lattice state on the square lattice.
Our work on nearest neighbor repulsive biquadratic interaction points %out 
towards new regimes to search for interesting behaviors in future numerical work.

{\bf Aknowdegements:}  We would like to acknowledge helpful discussions and suggestions from H-C. Jiang, P. Coleman, P. Chandra, Y. Jiang, F. Wang, and R. Tomalle.  SAK was supported, in part, by  NSF grant \# DMR-1608055 at Stanford.

\bibliography{citation}{}

\clearpage
\onecolumngrid
\appendix
\section{Holstein-Primakoff transformation for coplanar(including collinear) configuration}
In this section, we would like to review a systematic way to perform Holstein-Primakoff transformation, if the classical configuration is coplanar. This will eventually lead to a translational invariant Hamiltonian, in terms of creation and annihilation operators, for Neel, stripe and SVC state.
Without loss of generality, let us suppose spins in the classical configuration is in xz plane $\vec{n_i}=(\sin\theta_i,0,\cos\theta_i)$. 

If $\theta=0$, the standard Holstein-Primakoff transformation is 
\begin{equation}
\begin{split}
(J_{x0},J_{y0},J_{z0})=(\sqrt{2S}\frac{b+b^\dagger}{2},\sqrt{2S}\frac{b-b^\dagger}{2i},S-b^\dagger{b})
\end{split}
\end{equation}
Here, we have already taken the leading order contribution in HP transformation. 
Generally, for arbitrary $\theta_i$, we can choose
\begin{equation}
\begin{split}
&J_x=J_{x0}\cos\theta_i+J_{z0}\sin\theta_i\\
&J_y=J_{y0}\\
&J_z=J_{z0}\cos\theta_i-J_{x0}\sin\theta_i
\end{split}
\end{equation}
Now we would like to explicitly write down $\vec{J_i}\cdot\vec{J_j}$, up to quadratic terms in terms of bosonic operators. The constant term $\vec{J_i}\cdot\vec{J_j}$ is $S^2\cos\theta$ if $J_i$ and $J_j$ are different spin, and $S(S+1)$ if $i=j$. This will be important when calculating ground state energy.
The linear terms in $\vec{J_i}\cdot\vec{J_j}$ are
\begin{equation}
\begin{split}
\frac{1}{S}(\vec{J_i}\cdot\vec{J_j})_{\text{linear}}=\frac{\sin\theta}{\sqrt{2}}(b_i+b_i^\dagger-b_j-b_j^\dagger)
\end{split}
\end{equation}
, which is useful in the biquadratic interaction $(\vec{J_i}\cdot\vec{J_j})^2$, which contains the square of the above linear terms.
The quadratic terms in $\vec{J_i}\cdot\vec{J_j}$ is
\begin{equation}
\begin{split}
\frac{1}{S}(\vec{J_i}\cdot\vec{J_j})_{\text{quadratic}}=-(b_i^\dagger{b_i}+b_j^\dagger{b_j})\cos\theta+\frac{1}{2}(1+\cos\theta)(b_i^\dagger{b_j}+c.c)+\frac{1}{2}(-1+\cos\theta)(b_i^\dagger{b_j^\dagger}+c.c)
\end{split}
\label{PH}
\end{equation}
$\theta$ is the angle difference between $i$ and $j$ in the classical configuration. 

For the coplanar Neel, stripe and SVC phase, $\cos\theta$ and $\sin^2\theta$ for nearest neighbored and next-nearest neighbored spins is invariant under translation. Therefore, a translational invariant Hamiltonian in terms of creation and annihilation operators is expected.

In the next subsections, we will explicit derive the resulted Hamiltonian for Neel and stripe phase. We will include the result for SVC phase.

\subsection{square lattice-Neel \& stripe phase}
In this section, we consider the general ordering vector $\vec{Q}=(\pi,\theta),\theta=0,\pi$, which describes the Neel-stripe transition. For the nearest neighbored $J_1$ terms, angle difference is $\pm\pi$ along x-direction and $\pm\theta$ along y-direction. Plugging in the Holstein-Primakoff transformation in Eq.\ref{PH}, we get
\begin{equation}
\begin{split}
(J_1S)\times\left[\sum_{i}2(1-\cos\theta)b_i^\dagger{b_i}+\sum_{\langle{ij}\rangle{Y}}\frac{1}{2}(1+\cos\theta)(b_i^\dagger{b_j}+c.c.)+\sum_{\langle{ij}\rangle{Y}}\frac{1}{2}(-1+\cos\theta)(b_i^\dagger{b}_j^\dagger+c.c.)+\sum_{\langle{ij}\rangle{X}}(-1)(b_i^\dagger{b}_j^\dagger+c.c.)\right]
\end{split}
\end{equation}

Noted that the summation along x and y are separately written down. For second nearest neighbored $J_2$ terms, the angle difference is $\pm(\pi\pm\theta)$. After Holstein-Primakoff transformation, we have 
\begin{equation}
\begin{split}
(J_2S)\times\left[\sum_{i}4\cos{\theta}b_i^\dagger{b_i}+\sum_{\langle\langle{ij}\rangle\rangle}\frac{1}{2}(1-\cos\theta)(b_i^\dagger{b_j}+c.c.)+\sum_{\langle\langle{ij}\rangle\rangle}\frac{1}{2}(-1-\cos\theta)(b_i^\dagger{b_j}+c.c.)\right]
\end{split}
\end{equation}

For the biquadratic terms, we first perform mean-field approximation, and then Holstein-Primakoff transformation:
\begin{equation}
\begin{split}
&-K/S^2\sum_{\langle{ij}\rangle}(\vec{J_i}\cdot\vec{J_j})^2=2K\sum_{\langle{ij}\rangle_X}\vec{J_i}\cdot\vec{J_j}-2K\cos\theta\sum_{\langle{ij}\rangle_Y}\vec{J_i}\cdot\vec{J_j}-KNS^2(1+\cos^2\theta)\\
&=(2KS)\times\left[\sum_{i}2(1+\cos^2\theta)b_i^\dagger{b_i}+\sum_{\langle{ij}\rangle{Y}}\frac{1}{2}(-1-\cos\theta)\cos\theta(b_i^\dagger{b_j}+c.c.)\right.\\&\left.+\sum_{\langle{ij}\rangle{Y}}\frac{1}{2}(1-\cos\theta)\cos\theta(b_i^\dagger{b}_j^\dagger+c.c.)+\sum_{\langle{ij}\rangle{X}}(-1)(b_i^\dagger{b}_j^\dagger+c.c.)\right]-KNS^2(1+\cos^2\theta)
\end{split}
\end{equation}
N is the total number of sites. 

Now we can perform Fourier transformation 
\begin{equation}
\begin{split}
b_i^\dagger=\frac{1}{\sqrt{N}}\sum_{\vec{k}}e^{i\vec{k}\cdot\bf{R_i}}b_{\vec{k}}^\dagger
\end{split}
\end{equation}
We have
\begin{equation}
\begin{split}
&H=\sum_{\vec{k}}A_k(b_k^\dagger{b}_k+b_{-k}^\dagger{b}_{-k})+B_k(b_k^\dagger{b}_{-k}^\dagger+{b_k}{b}_{-k})\\
&A_k/S=[(1-\cos\theta)+\frac{1}{2}(\cos\theta+1)\cos{k_y}]J_1+[2\cos\theta+(1-\cos\theta)\cos{k_x}\cos{k_y}]J_2\\
&+[(1+\cos^2\theta)+\frac{1}{2}(-\cos\theta-1)\cos\theta\cos{k_y}](2K)\\
&B_k/S=[\frac{1}{2}(\cos\theta-1)\cos{k_y}-\cos{k_x}]J_1+(-1-\cos\theta)\cos{k_x}\cos{k_y}J_2\\
&+[\frac{1}{2}(-\cos\theta+1)\cos\theta\cos{k_y}-\cos{k_x}](2K)
\end{split}
\end{equation}
By taking $\theta=0,\pi$, we can reproduce the result for stripe and Neel phase, as obtained in \citep{Fang2008}.

\subsection{square lattice-SVC phase}
In the classical configuration for the SVC state, the angle between nearest neighbors is $\pi/2$, and the angle difference between next-nearest neighbors is $\pi$. The Hamiltonian in terms of bosonic operators is
\begin{equation}
\begin{split}
&H=\sum_{\vec{k}}A_k(b_k^\dagger{b}_k+b_{-k}^\dagger{b}_{-k})+B_k(b_k^\dagger{b}_{-k}^\dagger+{b_k}{b}_{-k})+const.\\
&A_k/S=\frac{J_1}{2}(\cos{k_x}+\cos{k_y})+2J_2-K(2-\cos{k_x}-\cos{k_y})\\
&B_k/S=-\frac{J_1}{2}(\cos{k_x}+\cos{k_y})-2J_2\cos{k_x}\cos{k_y}-K(2-\cos{k_x}-\cos{k_y})
\end{split}
\end{equation}

\subsection{triangular lattice-$120^\circ$ phase $\&$ stripe phase}
In this section, we will apply Holstein-Primakoff transformation to the coplanar $120^\circ$ phase and stripe phase. We will explicitly derive the $120^\circ$ phase, while provide the result for the stripe phase.

For $120^\circ$ phase, the angle difference is $120^\circ$ for all nearest neighbors. Plugging in the Holstein-Primakoff transformation, we get
\begin{equation}
\begin{split}
(J_1S)\times\left[\sum_{i}3b_i^\dagger{b_i}+\sum_{\langle{ij}\rangle}\frac{1}{4}(b_i^\dagger{b_j}+c.c.)+\sum_{\langle{ij}}-\frac{3}{4}(b_i^\dagger{b}_j^\dagger+c.c.)\right]
\end{split}
\end{equation}

For the next-nearest neighbored $J_2$ terms, angle difference is $0^\circ$, that leads to
\begin{equation}
\begin{split}
(J_2S)\times\left[\sum_{i}-6b_i^\dagger{b_i}+\sum_{\langle{ij}\rangle}(b_i^\dagger{b_j}+c.c.)\right]
\end{split}
\end{equation}

Biquadratic terms can be treated by performing mean-field approximation. Firstly, similar to the derivation of Neel phase for square lattice, it effectively modifies the nearest neighbor coupling constant $J_1$ as follows:
\begin{equation}
\begin{split}
J_1\rightarrow{J_1-2K/S^2\langle{J_i\cdot{J_j}}\rangle}=J_1+K
\end{split}
\end{equation}
Secondly, since the configuration is not collinear, there is an extra term from the square of linear term in HP transformation
\begin{equation}
\begin{split}
-K\sin^2{120^\circ}\sum_{\langle{ij}\rangle}(J_{i,x0}-J_{j,x0})^2
\end{split}
\end{equation}

After performing Fourier transformation, the final result for $120^\circ$ phase is
\begin{equation}
\begin{split}
&H=E_{cl}+\sum_{\vec{k}}A_k(b_k^\dagger{b}_k+b_{-k}^\dagger{b}_{-k})+B_k(b_k^\dagger{b}_{-k}^\dagger+{b_k}{b}_{-k})-\frac{9KSN}{8}\\
&A_k/S=(J_1+K)\left[\frac{3}{2}(1+\frac{1}{2}\gamma_k)\right]+J_2(-3+3\gamma_k')-\frac{9K}{4}(1-\gamma_k)\\
&B_k/S=(J_1+K)(-\frac{9}{4}\gamma_k)-\frac{9K}{4}(1-\gamma_k)\\
&\gamma_k=\frac{1}{6}\sum_{\vec{\delta_1}}\exp(i\vec{k}\cdot\vec{\delta_1})=\frac{1}{3}(\cos{k_y}+2\cos{\frac{\sqrt{3}k_x}{2}}\cos{\frac{k_y}{2}})\\
&\gamma_k'=\frac{1}{6}\sum_{\vec{\delta_2}}\exp(i\vec{k}\cdot\vec{\delta_2})=\frac{1}{3}(\cos{\sqrt{3}k_x}+2\cos{\frac{\sqrt{3}k_x}{2}}\cos{\frac{3k_y}{2}}).
\end{split}
\nonumber
\end{equation}

The final result for stripe phase is 
\begin{equation}
\begin{split}
&H=E_{cl}+\sum_{\vec{k}}A_k(b_k^\dagger{b}_k+b_{-k}^\dagger{b}_{-k})+B_k(b_k^\dagger{b}_{-k}^\dagger+{b_k}{b}_{-k})\\
&A_k/S=J_1(1+\cos{k_y})+J_2(1+\cos{\sqrt{3}k_x})+2K(3-\cos{k_y})\\
&B_k/S=-2J_1(\cos{\frac{\sqrt{3}k_x}{2}}\cos{\frac{k_y}{2}})-2J_2(\cos{\frac{\sqrt{3}k_x}{2}}\cos{\frac{3k_y}{2}})-4K(\cos{\frac{\sqrt{3}k_x}{2}}\cos{\frac{k_y}{2}})
\end{split}
\nonumber
\end{equation}

\section{Holstein-primakoff transformation for non-coplanar configuration}
In this section, we focus on the system with non-coplanar classical configuration. The spins in the classical configuration is in general $\vec{n}=(\sin\theta\cos\phi,\sin\theta\sin\phi,\cos\theta)$. 

If $\theta=0$, standard Holstein-Primakoff transformation for $S$ is 
\begin{equation}
\begin{split}
(J_{x0},J_{y0},J_{z0})=(\sqrt{2S}\frac{b+b^\dagger}{2},\sqrt{2S}\frac{b-b^\dagger}{2i},S-b^\dagger{b})
\end{split}
\end{equation}

Generally, for arbitrary $\theta$ and $\phi$, we can set
\begin{equation}
\begin{split}
&J_x=J_{x0}\cos\theta\cos\phi-J_{y0}\sin\phi+J_{z0}\sin\theta\cos\phi\\
&J_y=J_{x0}\cos\theta\sin\phi+J_{y0}\cos\phi+J_{z0}\sin\theta\sin\phi\\
&J_z=-J_{x0}\sin\theta+J_{z0}\cos\theta
\end{split}
\end{equation}
Noted that commutation relationship and classical extrapolation still holds if we perform the following transformation
\begin{equation}
\begin{split}
J_{x0}\rightarrow{J_{x0}\cos\alpha+J_{y0}\sin\alpha}\\
J_{y0}\rightarrow{J_{y0}\cos\alpha-J_{x0}\sin\alpha}
\end{split}
\end{equation}
, which is useful when simplifying the result.

\subsection{square lattice-CSVC phase}
For the classical configuration of CSVC state, the z-components of spins follow Neel ordering, while xy-components follow SVC ordering. Let the magnitude of the z-component be $\cos{\phi}$, so xy-component has magnitude of $\sin\phi$. By minimizing the classical Hamiltonian, we get 
\begin{equation}
\cos^2\phi=\frac{J_1-2J_2}{|2K|}
\end{equation}
Here we would like to explicitly write HP transformation for 4 spins in a square, which corresponds to four sublattices in the system. Spins in the same sublattice should be expressed in the same way.
\begin{equation}
\begin{split}
&J_i=(J_{z0}\sin\phi+J_{x0}\cos\phi,J_{y0},J_{z0}\cos\phi-J_{x0}\sin\phi)\\
&J_j=(J_{y0},J_{z0}\sin\phi+J_{x0}\cos\phi,-J_{z0}\cos\phi+J_{x0}\sin\phi)\\
&J_k=(-J_{y0},-J_{z0}\sin\phi-J_{x0}\cos\phi,-J_{z0}\cos\phi+J_{x0}\sin\phi)\\
&J_l=(-J_{z0}\sin\phi-J_{x0}\cos\phi,-J_{y0},J_{z0}\cos\phi-J_{x0}\sin\phi)
\end{split}
\end{equation}
Here $J_i$ and $J_l$ are on one diagonal, while $J_j$ and $J_k$ are on the other. The above HP transformation will produce the following translational invariant Hamiltonian, in terms of creation and annihilation operators.
\begin{equation}
\begin{split}
&H=\sum_{\vec{k}}A_k(b_k^\dagger{b}_k+b_{-k}^\dagger{b}_{-k})+B_k(b_k^\dagger{b}_{-k}^\dagger+{b_k}{b}_{-k})+const.\\
&A_k/S=(J_1+2K\cos^2\phi)\left[2\cos^2\phi-\frac{1}{2}\sin^2\phi(\cos{k_x}+\cos{k_y})\right]+J_2\left[2-4\cos^2\phi-2\cos^2\phi\cos{k_x}\cos{k_y}\right]\\
&-K\sin^2\phi(1+\cos^2\phi)(2+\cos{k_x}+\cos{k_y})\\
&B_k/S=(J_1+2K\cos^2\phi)\left[-\frac{1}{2}\sin^2\phi(\cos{k_x}+\cos{k_y})-i\cos\phi(\cos{k_x}-\cos{k_y})\right]+J_2\left[2\sin^2\phi\cos{k_x}\cos{k_y}\right]\\
&+K\left[\sin^4\phi(2+\cos{k_x}+\cos{k_y})+2i\sin^2\phi\cos\phi(\cos{k_x}-\cos{k_y})\right]
\end{split}
\end{equation}
One can check that the above $A_k$ and $B_k$ can recover spin wave excitation energy for Neel phase and vortex lattice phase by taking $\phi=0$ and $\phi=\pi/2$. Noted that the classical configuration has four sublattices, so when comparing spin wave excitation energies, we need to fold the Brillouin zone back to $k_x\in[-\pi/2,\pi/2], k_y\in[-\pi/2,\pi/2]$. 

\subsection{triangular lattice-STC phase}
For STC state, we decompose the system into four sublattice. HP transformation for spins in the four sublattices is

\begin{equation}
\begin{split}
&J_i=(x,y,z)\\
&J_j=(cc'x-cs'z+sy,-cy+sc'x-ss'z,-c'z-s'x)\\
&J_k=(-c'x+s'z,y,-c'z-s'x)\\
&J_l=(cc'x-cs'z-sy,-cy-sc'x+ss'z,-c'z-s'x)
\end{split}
\end{equation}
Here, $(x,y,z)$ are short for $(S_{x0},S_{y0},S_{z0})$. $c,s=(1/2,\sqrt{3}/2)$ and $(c',s')=(1/3,2\sqrt{2}/3)$.

The Hamiltonian is
\begin{equation}
\begin{split}
&H=\sum_k\Phi\left[\begin{array}{cc}
A_k & B_k\\
B_k^\dagger & A_{-k}^*
\end{array}\right]\Phi^\dagger-\frac{4K}{3}(4NS)\\
&\Phi=[b_{1k},\;b_{2k},\;b_{3k},\;b_{4k},\;b_{1-k}^\dagger,\;b_{2-k}^\dagger,\;b_{3-k}^\dagger,\;b_{4-k}^\dagger]\\
&A_k=(J_1+\frac{2}{3}K)A_J\circ{M_1}+J_2A_J\circ{M_2}-\frac{8K}{9}A_K\circ{M_1}\\
&B_k=(J_1+\frac{2}{3}K)B_J\circ{M_1}+J_2B_J\circ{M_2}-\frac{8K}{9}B_K\circ{M_1}
\end{split}
\end{equation}
Due to the limit of space, we express $A_k$ and $B_k$ using element-wise product $\circ$. The above $4\times{4}$ matrices are
\begin{equation}
\begin{split}
&A_J=\left[\begin{array}{cccc}
1 & -1/6+i\sqrt{3}/6&1/3&-1/6-i\sqrt{3}/6\\
-1/6-i\sqrt{3}/6&1&1/6+i\sqrt{3}/6&1/6-i\sqrt{3}/6\\
1/3&1/6-i\sqrt{3}/6&1&1/6+i\sqrt{3}/6\\
-1/6+i\sqrt{3}/6&1/6+i\sqrt{3}/6&1/6-i\sqrt{3}/6&1
\end{array}\right]\\
&A_K=\left[\begin{array}{cccc}
3 & 1/2-i\sqrt{3}/2&-1&1/2+i\sqrt{3}/2\\
1/2+i\sqrt{3}/2&3&-1/2-i\sqrt{3}/2&-1/2+i\sqrt{3}/2\\
-1&-1/2+i\sqrt{3}/2&3&-1/2-i\sqrt{3}/2\\
1/2-i\sqrt{3}/2&-1/2-i\sqrt{3}/2&-1/2+i\sqrt{3}/2&3
\end{array}\right]\\
&B_J=\left[\begin{array}{cccc}
0& 1/3-i\sqrt{3}/3&-2/3&1/3+i\sqrt{3}/3\\
1/3-i\sqrt{3}/3&0&2/3&2/3\\
-2/3&2/3&0&2/3\\
1/3+i\sqrt{3}/3&2/3&2/3&0
\end{array}\right]\\
&B_K=\left[\begin{array}{cccc}
0 & 1/2-i\sqrt{3}/2&-1&1/2+i\sqrt{3}/2\\
1/2-i\sqrt{3}/2&0&1&1\\
-1&1&0&1\\
1/2+i\sqrt{3}/2&1&1&0
\end{array}\right]\\
&M_1=\frac{1}{2}\left[\begin{array}{cccc}
2 &1+e^{-i\sqrt{3}k_x+ik_y}&1+e^{2ik_y}&1+e^{i\sqrt{3}k_x+ik_y}\\
1+e^{i\sqrt{3}k_x-ik_y}&2&1+e^{i\sqrt{3}k_x+ik_y}&e^{i\sqrt{3}k_x+ik_y}+e^{i\sqrt{3}k_x-ik_y}\\
1+e^{-2ik_y}&1+e^{-i\sqrt{3}k_x-ik_y}&2&1+e^{i\sqrt{3}k_x-ik_y}\\
1+e^{-i\sqrt{3}k_x-ik_y}&e^{-i\sqrt{3}k_x-ik_y}+e^{-i\sqrt{3}k_x+ik_y}&1+e^{-i\sqrt{3}k_x+ik_y}&2
\end{array}\right]\\
&M_2=\frac{1}{2}\left[\begin{array}{cccc}
2 &e^{-i\sqrt{3}k_x-ik_y}+e^{2ik_y}&e^{-i\sqrt{3}k_x+ik_y}+e^{i\sqrt{3}k_x+ik_y}&e^{i\sqrt{3}k_x-ik_y}+e^{2ik_y}\\
e^{i\sqrt{3}k_x+ik_y}+e^{-2ik_y}&2&e^{i\sqrt{3}k_x-ik_y}+e^{2ik_y}&1+e^{2i\sqrt{3}k_x}\\
e^{i\sqrt{3}k_x-ik_y}+e^{-i\sqrt{3}k_x-ik_y}&e^{-i\sqrt{3}k_x+ik_y}+e^{-2ik_y}&2&e^{i\sqrt{3}k_x+ik_y}+e^{-2ik_y}\\
e^{-i\sqrt{3}k_x+ik_y}+e^{-2ik_y}&1+e^{-2i\sqrt{3}k_x}&e^{-i\sqrt{3}k_x-ik_y}+e^{2ik_y}&2
\end{array}\right].
\end{split}
\end{equation}
Hamiltonian can then be diagonalized using standard Bogoliubov transformation.

\section{square lattice-nematic \& SNVC \& chiral phase}
Let us first consider SNVC phase. Without loss of generality, let us consider $S_i$ along x and $S_{i+\hat{x}}$ along z. The cross-product is then $(YZ+XY,-XX-ZZ,ZY-YX)$. Here letter $X$ is short for $S_{x0}$, and we keep the first letter for $S_i$, and second letter for $S_{i+\hat{x}}$. Now we can plug in HP transformation. After summing over site i, only the second component of cross product is non-zero, since x and z components contain integration of odd function of $k_x$. The result is
\begin{equation}
\begin{split}
&\mathcal{S}_{SNVC}=\sum_{\vec{k}}(2-\frac{\cos(k_x)+\cos(k_y)}{2})v_k^2+(-\frac{\cos(k_x)+\cos(k_y)}{2})u_kv_k\\
&v_k^2=\frac{1}{2}(\frac{|A_k|}{\sqrt{A_k^2-B_k^2}}-1)\\
&u_kv_k=-\frac{1}{2}\frac{B_k}{\sqrt{A_k^2-B_k^2}}\text{sign}(A_k)
\end{split}
\end{equation}

For chiral state, we can perform the same calculation. We first write order parameter in terms of $S_{x0,y0,z0}$, and then plug in HP transformation. Eventually we sum up all sites. The result is
\begin{equation}
\begin{split}
\mathcal{S}_{chir}=\frac{1}{\cos\phi}\sum_{\vec{k}}\left[3\cos\phi-\cos\phi\cos(k_x-k_y)+\frac{1}{2}(\cos{k_x}+\cos{k_y})\right]v_k^2+\left[\cos\phi\cos(k_x-k_y)+\frac{1}{2}(\cos{k_x}+\cos{k_y})\right]u_kv_k
\end{split}
\end{equation}

\end{document}